\documentclass[preprint,showpacs,preprintnumbers,amsmath,amssymb,12pt]{revtex4-1}
\usepackage{lipsum}
\usepackage{setspace}
\usepackage{bm}
\usepackage[colorlinks,linkcolor = black,anchorcolor = blue,urlcolor = blue,	citecolor = blue]{hyperref}
\usepackage{float}
\usepackage{graphicx}
\usepackage{makecell}
\usepackage[utf8]{inputenc}

\begin{document}
\title{Mott-Derived Local Moments and Kondo Hybridization in a $d$-Electron Kagome Lattice}

\author{Xing Zhang$^{1,2,\dagger}$, Xintong Li$^{1,2\dagger\bigstar}$, Boqin Song$^{1,2,\dagger}$, Yuyang Xie$^{1,2,\dagger}$, Qinghong Wang$^{1,2,\dagger}$, Taimin Miao$^{1,2}$, Shusen Ye$^{3}$, Junhao Liu$^{1,2}$, Bo Liang$^{1,2}$, Neng Cai$^{1,2}$, Hao Chen$^{1,2}$, Wenpei Zhu$^{1,2}$, Mingkai Xu$^{1,2}$, Wei-Jian Li$^{4}$, Shun-Li Yu$^{4}$, Shenjin Zhang$^{5}$, Fengfeng Zhang$^{5}$, Feng Yang$^{5}$, Zhimin Wang$^{5}$, Qinjun Peng$^{5}$, Hanqing Mao$^{1,2\bigstar}$, Zhihai Zhu$^{1,2}$, Guodong Liu$^{1,2}$, Zuyan Xu$^{5}$, Yi-feng Yang$^{1,2}$, Tianping Ying$^{1,2\bigstar}$, Lin Zhao$^{1,2\bigstar}$ and X. J. Zhou$^{1,2\bigstar}$}

\affiliation{
\\$^{1}$Beijing National Laboratory for Condensed Matter Physics, Institute of Physics, Chinese Academy of Sciences, Beijing 100190, China.
\\$^{2}$University of Chinese Academy of Sciences, Beijing 100049, China.
\\$^{3}$State Key Laboratory of Low Dimensional Quantum Physics, Department of Physics, Tsinghua University, Beijing 100084, China.
\\$^{4}$National Laboratory of Solid State Microstructures and Department of Physics, Nanjing University, Nanjing 210093, China.
\\$^{5}$Technical Institute of Physics and Chemistry, Chinese Academy of Sciences, Beijing 100190, China.
}

\date{\today}
\maketitle
\noindent{\bf 
Unlike canonical Kondo lattices in $f$-electron systems, where localized $f$ orbitals naturally provide local moments, $d$-electron Kondo lattices require a distinct mechanism for local-moment formation. However, the study of $d$-electron Kondo lattices in bulk materials remains far from settled, particularly with regard to the microscopic origin of the local moments. Here, we report a microscopic mechanism for this process in the bilayer kagome metal CsCr$_6$Sb$_6$, where strong correlations drive a Mott splitting of the kagome flat band to supply the requisite local moments. By
combining scanning tunneling microscopy/spectroscopy (STM/STS) and angle-resolved photoemission spectroscopy (ARPES), we resolve a spectroscopic hierarchy between high-energy correlation effects and low-temperature hybridization. Low-temperature STS reveals a robust asymmetric suppression of the density of states near $E_{\mathrm{F}}$ that is well captured phenomenologically by a Fano-type lineshape, while ARPES detects a sharp quasiparticle peak near $E_{\mathrm{F}}$ at the $M$ point and near $K$. These low-energy signatures evolve on the same temperature scale and disappear upon warming, consistent with the onset of Kondo hybridization. At the same time, STS resolves symmetric humps at approximately $\pm 50$ mV and ARPES identifies a weakly dispersive feature around 50 meV below $E_{\mathrm{F}}$; unlike the near-$E_{\mathrm{F}}$ hybridization signatures, these features persist to substantially higher temperatures. This separation of energy and temperature scales supports a two-stage picture in which a kagome flat band first undergoes correlation-driven splitting into lower and upper Hubbard bands, and the occupied lower Hubbard band supplies the local moments that later hybridize with itinerant electrons at lower temperature. Our results therefore move beyond the phenomenology of a kagome Kondo lattice candidate and instead provide a microscopic spectroscopic picture linking Mottness to Kondo hybridization in a frustrated $d$-electron system.
}

\vspace{3mm}
\section{INTRODUCTION}
The Kondo lattice is one of the central paradigms of strongly correlated electron physics. It arises from the coupling between a periodic array of localized magnetic moments and itinerant electrons, and underlies heavy-fermion behavior, unconventional quantum criticality, and a wide range of competing ordered states\cite{coleman2015heavyfermionskondolattice,ReviewRMP2020,HeavyFermion2019}. In conventional materials this physics is most naturally realized in $f$-electron compounds, where the strong spatial localization of $4f$ or $5f$ orbitals provides robust local moments\cite{RajS2005,Fujimori2006,KummerK2015,ParkWK2012}. By contrast, realizing coherent Kondo-lattice behavior in bulk $d$-electron materials is considerably more challenging because $d$ electrons are typically more itinerant and therefore less likely to furnish well-defined local moments.

Recent progress has broadened this landscape. Artificially engineered heterostructures and moir\'e superlattices have shown that Kondo lattice behavior can emerge beyond traditional $f$-electron materials once localized and itinerant electronic sectors coexist\cite{MoireKondo2023,
vanderwaalsKondo2024,Vano2021Nature}. In layered heterostructures such as 1T-TaS$_2$/1H-TaS$_2$, localized moments residing in an insulating layer can couple to itinerant electrons in an adjacent metallic layer\cite{Vano2021Nature}. Similarly, in moir\'e superlattices formed by stacking two-dimensional materials, the formation of narrow or flat bands can give rise to strongly localized electronic states, effectively supplying localized magnetic moments\cite{MoireKondo2023}. Geometrically frustrated lattices provide an especially appealing intrinsic route because their topology can generate flat bands with strongly suppressed kinetic energy. Such flat bands may provide localized moments that participate in Kondo hybridization, offering an alternative pathway to Kondo lattice behavior in non-$f$-electron systems\cite{DurgaK2020SA,LiVoKondoNC2012,Ni3InKondoNP2024,ZahidM2020PRL,Checkelsky2024NRM,WangYiLin2025PRB}. In this setting, the key unresolved issue is not simply whether Kondo lattice behavior can occur, but how the local moments required for Kondo hybridization are generated microscopically in a native $d$-electron material.

The kagome lattice is a particularly promising platform for addressing this question. Its electronic structure naturally hosts Dirac cones, van Hove singularities, and a flat band originating from destructive quantum interference\cite{Syozi1951,JXY2022Nature}. If such a flat band lies close to $E_{\mathrm{F}}$ and is sufficiently correlated, it may become the source of local moments that subsequently hybridize with itinerant carriers\cite{ZahidM2020PRL,Wangyaojia2023review,Ni3InKondoNP2024,WangYiLin2025PRB}. This route is conceptually distinct from conventional $f$-electron Kondo lattices, because the local moments would not be atomic in origin but instead emerge from the interplay between lattice geometry and electron correlation.

CsCr$_6$Sb$_6$, a bilayer kagome material with low-energy states derived predominantly from Cr 3$d$ orbitals, has recently been established as a kagome Kondo lattice candidate from bulk transport, thermodynamic, and spectroscopic measurements\cite{Songboqin2025NC}. That work demonstrated the realization of Kondo-lattice behavior primarily through transport measurements and highlighted the unusual coexistence of flat bands, ultra-low carrier density, and dimensionality-tuned many-body effects. What remained unresolved, however, was the microscopic origin of the local moments and the relation between the flat-band correlation physics and the low-temperature hybridized state.

Here we focus on that unresolved issue. By combining STM/STS with ARPES, we identify two spectroscopically distinct components of the low-energy electronic structure in CsCr$_6$Sb$_6$. The first is a near-$E_{\mathrm{F}}$ hybridization feature that develops only at low temperature and disappears on warming. The second is a pair of high-energy features near $\pm 50$ meV that remain robust to much higher temperature. The coexistence but distinct thermal evolution of these features leads us to a more specific microscopic picture: a kagome flat band first undergoes correlation-driven splitting into Hubbard-like states, and the occupied lower Hubbard band then provides the local moments that participate in low-temperature Kondo hybridization. In this way, the present work is aimed not at re-establishing the existence of a kagome Kondo lattice in CsCr$_6$Sb$_6$, but at resolving its spectroscopic hierarchy and clarifying the likely Mott-derived origin of the local moments.

\vspace{3mm}
\section{Crystal structure and transport context}
CsCr$_6$Sb$_6$ crystallizes in space group $R\bar{3}m$ with the lattice parameters a = 5.546 \text{Å} and c = 34.52 \text{Å}\cite{Songboqin2025NC}. This system realizes a double kagome (DK) lattice structure. As illustrated in Fig.~\ref{Fig1}(a), the crystal structure consists of a bilayer (Cr$_3$Sb)$_2$ stacking network, which is sandwiched between two honeycomb Sb2 layers and separated by hexagonally arranged Cs atoms. A prototypical Cr$_3$Sb plane is composed of the Cr kagome sublattice and a Sb1 hexagonal sublattice. 
This stacking results in a van der Waals–like layered structure, giving rise to two natural cleavage planes: the Cs-terminated surface and the Sb2-terminated surface (Fig.~\ref{Fig1}(b) and Fig.~\ref{Fig1}(c)).
Fig.~\ref{Fig1}(d) shows the temperature dependence of the in-plane resistivity of CsCr$_6$Sb$_6$.
An upturn in resistivity emerges below approximately 85 K, consistent with previous reports\cite{Songboqin2025NC} and is commonly attributed to enhanced scattering between localized magnetic moments and itinerant electrons.
The first derivative d$R$/d$T$ (inset of Fig.~\ref{Fig1}(d)) reveals a kink at $T_{\mathrm{F}} \sim 75$~K, indicative of a frustrated magnetic transition. This behavior is further supported by magnetic susceptibility measurements reported previously.
Consistent with these observations, $\mu$SR measurements have demonstrated that the magnetism in bulk CsCr$_6$Sb$_6$ corresponds to a short-ranged magnetic order\cite{XiangqiLiu2025arxiv}.

\vspace{3mm}
\section{STM/STS results}
Fig.~\ref{Fig2}(a) and \ref{Fig2}(b) show low-temperature STM topographies acquired at approximately 5 K on the honeycomb Sb2-terminated surface and the reconstructed Cs-terminated surface, respectively. 
The Sb2-terminated surface, obtained by cleaving the crystal at around 20 K, exhibits a clean honeycomb lattice of Sb atoms with randomly distributed residual Cs atoms appearing as bright protrusions. 
In contrast to several kagome materials that host charge density waves (CDW)\cite{CVScdw_PRB2021,CVS_Nature2021,
CVScdw_PRX2021,Kundu2024}, no CDW modulation is observed here, as illustrated by the bare honeycomb lattice of Sb atoms (inset).
The Cs-terminated surface (Fig.~\ref{Fig2}(b)), obtained by room-temperature cleavage, displays more complex surface reconstructions arising from partial desorption of Cs atoms. Two representative reconstruction patterns are observed: a well-ordered hexagonal lattice and a maze-like configuration composed of short-range disordered stripes.
According to the corresponding Fourier transform (FT) map (inset) and the temperature-dependent reconstructions in similar V-based kagome materials\cite{RbV3Sb5nanolett2022}, the periodic superstructure in the green circle is attributed to Cs-$\sqrt{3} \times \sqrt{3}$, indicating the 1/3 coverage of Cs-1×1 plane. 
On the other hand, the maze-like feature highlighted by the blue ellipse is formed by disordered short-range stripes with the periodicity of $\sqrt{3} \times 1$, representing the half coverage of Cs atoms. 
As the cleavage temperature decreases, the Cs-$\sqrt{3} \times \sqrt{3}$ reconstruction disappears and the short-range stripes are arranged into long-range unidirectional stripes (Supplementary Fig.~\ref{SI_Fig2}(b)), indicating that Cs-termination is dominated by Cs-$\sqrt{3} \times 1$ reconstruction. The Cs reconstructions are unchanged as the temperature rises, as shown in Fig.~\ref{SI_Fig5}.

Fig.~\ref{Fig2}(c) and Fig.~\ref{Fig2}(d) present the corresponding STS spectra acquired on both surfaces at $T$ = 5 K. Despite the distinctly different surface morphologies, the d$I$/d$V$ spectra are surprisingly similar in several key aspects. The spatially averaged STS spectrum on the Sb2 surface (Fig.~\ref{Fig2}(c)) exhibits a well-defined and ubiquitous gap-like feature with density of states (DOS) suppression around $E_{\mathrm{F}}$, which is also manifested on the Cs-termination (Fig.~\ref{Fig2}(d)), despite the presence of an additional positive-bias background contribution. 
Both spectra exhibit an asymmetric gap with substantial residual DOS at $E_{\mathrm{F}}$, reminiscent of the Fano resonance signatures commonly observed in Kondo lattice systems\cite{Fano1961PR,Madhavan1998Science,Figgins2010,SchmidtNature2010,ParkWK2012}.
Additionally, the spectra are virtually independent of the specific Cs reconstruction, as indicated by the green and blue curves in Fig.~\ref{Fig2}(d), with only subtle variations in detail. To clarify the resistivity behavior in Fig.~\ref{Fig1}(d), we performed variable-temperature STS measurements on both Sb2- and Cs-terminated surfaces. As shown in Fig.~\ref{Fig2}(e) and Fig.~\ref{Fig2}(f), both surfaces exhibit a similar evolution of the electronic structure: upon warming from 5 K, the DOS suppression near $E_{\mathrm{F}}$ is filled continuously and the gap closes at approximately 57 K. The dashed-line simulations in Fig.~\ref{Fig2}(e), obtained by convolving the 5 K spectrum with the derivative of the Fermi-Dirac distribution at each temperature, indicate that simple thermal broadening cannot account for the observed spectral evolution, supporting an intrinsic origin of the gap closing. 
In contrast, the peak near −13 mV persists to much higher temperatures, indicating a different origin that is most likely associated with a temperature-independent electronic band structure. This interpretation is further supported by our ARPES measurements, as discussed in the following section.

From the temperature evolution of the STS spectra, the low-temperature electronic structure can be resolved into two components: a temperature-dependent asymmetric gap-like structure and a temperature-independent single-peak feature. In light of previous transport results\cite{Songboqin2025NC}, the asymmetric gap-like structure can be naturally interpreted within a Fano-resonance framework, as commonly discussed in Kondo lattice systems, arising from tunneling interference between itinerant conduction states and localized electronic states\cite{SchmidtNature2010,ParkWK2012,Ruanwei2014PRL,Zhangyun2018PRB}. The corresponding spectral lineshape is well captured by the Fano formalism\cite{Fano1961PR}. In contrast, the temperature-independent peak exhibits a lineshape that can be approximated by a Gaussian function, distinguishing it from the Fano-type contribution. Accordingly, as shown in Fig.~\ref{Fig2}(g) and Fig.~\ref{Fig2}(h), the d$I$/d$V$ spectra acquired at 5 K on both surfaces are fitted using a combined Fano-plus-Gaussian model\cite{YunZhang2018}, enabling a quantitative analysis of the low-energy electronic structure relevant to Kondo lattice behavior in CsCr$_6$Sb$_6$. The Fano lineshape is given by:                                    
\begin{equation}
\frac{dI}{dV} \propto \frac{(\epsilon + q)^2}{1 + \epsilon^2}, 
\qquad 
\epsilon = \frac{eV - \epsilon_0}{\Gamma}
\label{eq:Fano}
\end{equation}
where the parameter $q$ represents the relative tunneling ratio into the localized and itinerant electronic states, taking values of approximately 0.2 on the Cs surface and −0.4 on the Sb2 surface. In Kondo lattice systems, the value of $q$ is known to be highly sensitive to surface termination and tunneling conditions\cite{vanderwaalsKondo2024}. 
When the cleaved plane predominantly hosts localized states, a characteristic spectral signature is the Kondo resonance, which manifests as a pronounced peak in the vicinity of $E_{\mathrm{F}}$\cite{Vano2021Nature,vanderwaalsKondo2024,ZahidM2020PRL,YunZhang2018}. 
Conversely, on a surface where itinerant electronic states dominate the tunneling process, the spectrum exhibits a gap-like feature, indicative of a suppression of the conduction-electron DOS due to Kondo screening\cite{Vano2021Nature,ParkWK2012,vanderwaalsKondo2024,Ernst2011Nature,Aynajian2012Nature,Ruanwei2014PRL}.
In CsCr$_6$Sb$_6$, the localized states are derived from the Cr 3$d$ electrons, which are responsible for the gap-like spectral features observed on both the Cs- and Sb2-terminated surfaces. The parameter $\epsilon_0$ denotes the energy location of the resonance, which is $\sim 2$~meV on Cs surface and $\sim -4$~meV on Sb2 surface. Even when the resonance features are located at comparable energies near $E_{\mathrm{F}}$, variations in the background electronic structure can lead to noticeable differences in the fitted parameters. The parameter $\Gamma$ corresponds to the half-width at half-maximum (HWHM) of the Fano function and provides an effective energy scale associated with Kondo hybridization. Within the single-impurity Kondo framework\cite{SingleKondoImpurity2002PRL}, this energy scale yields an estimated characteristic temperature of $T_{\mathrm{K\_onset}}$ = 53 K, which we associate with the onset of Kondo hybridization in  our system (Fig.~\ref{SI_Fig6}).

Strikingly, the STS spectra on both surfaces share another pronounced feature: a pair of humps at approximately $\pm 50$ mV, highlighted by purple shading in Fig.~\ref{Fig2}(c) and Fig.~\ref{Fig2}(d). In contrast to the gap-like feature, which disappears above 50 K, these hump features persist to significantly higher temperatures. They are reproducibly observed across multiple measurements performed on different STM systems (see Methods), underscoring their intrinsic nature.
Similar hump-like features have been reported in recent STM studies of other kagome systems, including Co$_{{1-}{x}}$Fe$_x$Sn and 2D metal-organic framework. In those systems, the humps were attributed to upper Hubbard band (UHB) and lower Hubbard band (LHB) arising from correlation-driven Mott splitting of a kagome flat band\cite{2024arXivBerthold,MottinMOF2025NSR}. As a kagome metal, CsCr$_6$Sb$_6$ is therefore expected to host analogous correlation effects, and the observed symmetric humps at $\pm 50$ mV are likely to share a similar origin, reminiscent of UHB and LHB. Further discussion of the origin of Mottness in CsCr$_6$Sb$_6$ is provided below in conjunction with the ARPES results.

\vspace{3mm}
\section{ARPES results}
To investigate the electronic signatures  of  Kondo hybridization in momentum space, we performed laser-based ARPES on CsCr$_6$Sb$_6$.
Figure \ref{Fig3} exhibits the measured  Fermi surface and band dispersions  along representative  high-symmetry directions of the Brillouin zone. The Fermi surface of CsCr$_6$Sb$_6$ shown in Fig.~\ref{Fig3}(a) is obtained by integrating the ARPES spectral weight within an energy window of $\pm 10$ meV around $E_{\mathrm{F}}$. The data are symmetrized assuming threefold rotational symmetry of the lattice  (measured at 16 K with sample bias of -69 V, see methods).
A small electron-like Fermi pocket is observed at the Brillouin-zone center ($\Gamma$  point). Around the zone corner (K point), the Fermi surface exhibits characteristic triangular pockets. Notably, a pronounced elliptical feature with strong spectral weight emerges at the zone boundary (M point), connecting adjacent triangular pockets around neighboring K points. The enhanced spectral weight near the M point indicates its substantial contribution to the low-energy electronic states. 
The overall Fermi surface topology near the Brillouin-zone boundary, schematically indicated by the blue contours in Fig.~\ref{Fig3}(a), closely resembles that reported for the kagome metal CsV$_3$Sb$_5$\cite{HuYong2022NC,
HuYong2023ARPES135}.

The corresponding band dispersions along high-symmetry directions, extracted from the raw data in Fig.~\ref{Fig3}(a), are shown in Fig.~\ref{Fig3}(b)-Fig.~\ref{Fig3}(d). Four distinct low-energy bands (labeled $\alpha, \beta, \varepsilon, \gamma$) can be resolved, as highlighted by black dotted lines and triangular markers.
The $\alpha$ band forms an electron pocket centered at the $\Gamma$ point, with its band bottom located approximately 10-15 meV below $E_{\mathrm{F}}$, consistent with the small Fermi pocket observed in Fig.~\ref{Fig3}(a). 
The $\beta$ band (marked by triangular symbols in Fig.~\ref{Fig3}(b)-(d)) appears as a weakly dispersive and relatively broad feature situated about 50 meV below $E_{\mathrm{F}}$ along all three high-symmetry cuts, consistent with a flat-band-derived electronic state. This band is more clearly resolved in ARPES measurements using helium-lamp and synchrotron-radiation light sources\cite{Songboqin2025NC,
XiangqiLiu2025arxiv}. In light of the preceding STS results, which reveal symmetric hump features consistent with Hubbard-like bands, this flat-band-derived feature is more likely associated with LHB rather than a conventional noninteracting kagome flat band. 
The $\varepsilon$ band remains largely non-dispersive and nearly indistinguishable from the $\beta$ band over most of momentum space. Near the K point, however, it exhibits a pronounced dispersion and crosses $E_{\mathrm{F}}$, giving rise to the triangular Fermi pockets observed around the zone corners.
All observed low-energy bands are confined within approximately 100 meV below $E_{\mathrm{F}}$ and are well separated from deeper valence bands by a gap of about 100–300 meV. As a result, the low-energy electronic properties are predominantly governed by these energetically isolated bands close to $E_{\mathrm{F}}$. According to the corresponding density functional theory (DFT) calculations\cite{Songboqin2025NC}, these bands originate almost exclusively from Cr 3$d$ orbitals. While the calculated band structures qualitatively capture the clustering of conduction bands near $E_{\mathrm{F}}$ and their separation from deeper valence bands by a sizable gap, quantitative differences remain, especially near the K point. Such deviations from DFT predictions suggest that electron correlation effects, which are not fully accounted for within standard DFT, play an important role in shaping the low-energy electronic structure.  

Fig.~\ref{Fig3}(e) shows the temperature-dependent band structure at the \text{M} point along the \text{M}-\text{K} direction (zero bias,  h$\nu$ = 6.994\, eV). 
At low temperature, two prominent features are observed: a weakly dispersive flat band located approximately 50 meV below $E_{\mathrm{F}}$ ($\beta$ band), and a pronounced spectral weight just below $E_{\mathrm{F}}$ at the M point.
With increasing temperature, the flat band remains nearly unchanged, whereas the strong spectral weight near $E_{\mathrm{F}}$ is progressively suppressed and eventually disappears at higher temperature(>50 K). 
To quantify this temperature evolution, the energy distribution curves (EDCs) were extracted by integrating the spectral weight around the M point at the momentum position indicated by the green arrow in Fig.~\ref{Fig3}(e). The resulting EDCs at different temperatures are shown in Fig.~\ref{Fig3}(g).
At low temperature, a sharp quasiparticle peak is clearly  observed near $E_{\mathrm{F}}$ as marked by the green arrow. This peak remains well defined up to 38 K, but rapidly weakens at higher temperatures and vanishes by 65 K. In contrast, the hump persists up to significantly higher temperatures, remaining visible above 95 K. 
Fig.~\ref{Fig3}(f) shows the low temperature band dispersion  along $\Gamma$-K direction (16 K, sample bias -49 V). 
Two bands crossing $E_{\mathrm{F}}$ are found: one in the vicinity of the K point ($\varepsilon$ band) and the other around the $\Gamma$ point ($\alpha$ band). 
The corresponding EDCs of the momentum positions indicated by the black and red arrows, are presented in Fig.~\ref{Fig3}(h) and Fig.~\ref{Fig3}(i), respectively. The temperature evolution of the band structure is presented in Fig.~\ref{SI_Fig8}.
A comparison of the quasiparticle spectral weights extracted from the EDCs fitting at three characteristic Fermi momenta reveals distinctly different temperature dependencies, as summarized in Fig.~\ref{Fig3}(j). The quasiparticle weight at the M point decreases continuously with increasing temperature and vanishes by 65 K. Although substantially weaker in overall intensity, the quasiparticle spectral weight near the K point exhibits a similar temperature dependence. In contrast, the quasiparticle spectral weight at the $\Gamma$ point remains nearly temperature independent and retains a strong intensity over the entire measured temperature range. This pronounced contrast indicates that the quasiparticles near the K point and at the M point exhibit fundamentally different thermal evolution compared to those at the $\Gamma$ point.
The temperature-insensitive behavior of the quasiparticle spectral weight at the $\Gamma$ point naturally explains why the peak at -13 meV observed in the STS spectrum (Fig.~\ref{Fig2}(e) and Fig.~\ref{Fig2}(f)) persists to much higher temperatures. Notably, the temperature at which the quasiparticle peaks vanish at the M point and near the K point coincides with the characteristic temperature $T_{\mathrm{K\_onset}}$ determined from STM measurements, indicating a common underlying origin. 
By combining the temperature evolution  of the STS gap (Fig.~\ref{Fig2}(f)) with that of the ARPES quasiparticle peak (Fig.~\ref{Fig3}(g) and Fig.~\ref{Fig3}(j)), we estimate $T_{\mathrm{K\_onset}}$ to lie in the range of 50-65 K, consistent with the value of $\sim 53$~K discussed above.
This temperature scale is significantly lower than the short-range magnetic ordering temperature $T_{\mathrm{F}} \sim 75$~K determined by transport measurements \cite{Songboqin2025NC,XiangqiLiu2025arxiv}. The separation between these two characteristic temperatures suggests that the observed temperature-dependent spectroscopic features are unlikely to be directly associated with the frustrated magnetic state. 

\vspace{3mm}
\section{Discussion and Conclusion}
Previous studies have reported no evidence for long-range magnetic order or other forms of long-range order below $T_{\mathrm{K\_onset}}$. Nevertheless, signatures suggestive of Kondo hybridization behavior at low temperatures have been inferred from transport and spectroscopic measurements\cite{Songboqin2025NC,XiangqiLiu2025arxiv}. Our low-temperature STM/STS measurements reveal a characteristic asymmetric Fano-type lineshape, which is commonly associated with Kondo resonance in Kondo lattice systems. These observations provide strong support for the presence of Kondo hybridization in CsCr$_6$Sb$_6$. In this context, it is natural to associate the temperature-dependent quasiparticle peaks observed at the M point and near the K point in our ARPES measurements with Kondo-induced band hybridization. 
In momentum-resolved ARPES measurements, Kondo hybridization near $E_{\mathrm{F}}$ is typically manifested by the emergence of coherent quasiparticle peak, a hallmark of Kondo lattice behavior that has been widely documented in heavy-fermion systems\cite{Zhangyun2018PRB,BookHeavyFermions1993,Zhangyun2020ReviewEng,XiangqiLiu2025arxiv,Zhangyun2022PRB,ChenQiuYun2017PRB,Chenqiuyun2019PRL,ZhangW2018PRB,Chenqiuyun2018PRL}.
It is also noteworthy that the robust and ubiquitous temperature-independent “$\pm 50$ mV” humps repeatedly observed in STM measurements across multiple samples highlight their intrinsic origin (see supplementary materials). In comparison with the ARPES data, the hump located at approximately “–50 mV” can be naturally associated with the $\beta$ band, which we identify as the LHB situated around 50 meV below $E_{\mathrm{F}}$.

Taken together, the STM/STS and ARPES results point to a hierarchy of scales rather than to a single undifferentiated correlation effect. The features that evolve together on the lower temperature scale are the asymmetric near -$E_{\mathrm{F}}$ spectral suppression in STS and the sharp quasiparticle peak in ARPES. The features that remain stable on a higher scale are the symmetric $\pm 50$~mV humps in STS and the weakly dispersive state around $-50$~meV in ARPES. Because the two classes of features exhibit clearly different thermal evolution, assigning them to the same underlying process would be unnatural.

Both STM/STS and ARPES therefore provide mutually consistent spectroscopic signatures indicative of Kondo hybridization in CsCr$_6$Sb$_6$. Unlike conventional Kondo lattice systems, however, this material does not host localized $f$ electrons, which raises the central question of the microscopic origin of the localized magnetic moments required for Kondo-lattice formation. Theoretically, electrons associated with kagome flat bands have been proposed as a potential source of localized spins, effectively playing a role analogous to that of localized $f$ electrons in conventional Kondo-lattice systems\cite{WangYiLin2025PRB}. Consistent with this scenario, DFT calculations of CsCr$_6$Sb$_6$ indicate that the low-energy electronic states near $E_{\mathrm{F}}$ are predominantly derived from Cr 3$d$ orbitals and include a kagome-derived flat band located in close proximity to $E_{\mathrm{F}}$\cite{Songboqin2025NC}. This flat band is expected to be close to half filling. In correlated electron systems, half-filled narrow bands near $E_{\mathrm{F}}$ are particularly susceptible to on-site Coulomb repulsion, frequently leading to correlation-driven band splitting and the formation of Hubbard bands\cite{ShenZhiXunRMP2003,Nb3Cl8MottPRX2023,Brandow1977,MITRMP1998}.
In this context, we associate the robust and comparatively temperature-insensitive $\pm 50$~mV humps observed in the STS spectra of Fig.~\ref{Fig2}(c) and Fig.~\ref{Fig2}(d) with the upper and lower Hubbard bands arising from correlation-induced splitting of the kagome flat band, as schematically illustrated in Fig.~\ref{Fig4}(a) and Fig.~\ref{Fig4}(b). The occupied lower-Hubbard-band-like feature at approximately $-50$~meV is detected by both STM/STS and ARPES, whereas the unoccupied upper-Hubbard-band-like feature at approximately $+50$~meV is accessible only in STM/STS. Crucially, the formation of the occupied lower Hubbard band provides a natural source of localized magnetic moments in this system. The localized Cr 3$d$ electrons associated with this occupied flat-band-derived state can then act as local moments that are screened by itinerant conduction electrons, giving rise to a Kondo-like resonance near $E_{\mathrm{F}}$ at low temperature, as illustrated schematically in Fig.~\ref{Fig4}(c). Although a direct probe of the local magnetic moments associated with the lower Hubbard band is still lacking, the combined STM/STS and ARPES results provide strong and self-consistent support for this interpretation.

ARPES measurements further elucidate the evolution of the hybridized electronic structure in CsCr$_6$Sb$_6$, as summarized schematically in Fig.~\ref{Fig4}(d) and Fig.~\ref{Fig4}(e). At temperatures well above $T_{\mathrm{K\_onset}}$, the band structure along high-symmetry directions can be resolved more directly: the $\beta$ band, identified here with the occupied lower-Hubbard-band-like state, appears as a broadened and weakly dispersive flat feature; the $\alpha$ band forms a small electron pocket around the $\Gamma$ point; the $\gamma$ band corresponds to an electron-like band with its bottom just above $E_{\mathrm{F}}$ near the $M$ point; and the $\varepsilon$ band crosses $E_{\mathrm{F}}$ near the $K$ point, giving rise to a hole-like pocket. Upon cooling below $T_{\mathrm{K\_onset}}$, Kondo hybridization sets in. The localized $d$ electrons associated with the $\beta$ band hybridize with itinerant $d$ electrons from the dispersive conduction bands near the renormalized Kondo-resonance energy $\varepsilon_0$, which lies very close to $E_{\mathrm{F}}$, as indicated by the dashed green line in Fig.~\ref{Fig4}(e).

Our ARPES data indicate that this hybridization predominantly involves electronic states near the $M$ point and near the $K$ point, while the $\alpha$ band near $\Gamma$ remains largely unaffected. This behavior points to a pronounced momentum dependence of the hybridization process\cite{Momentum2024CeSiI,Giannakis2019OrbitalSelectiveKondo}. As illustrated schematically in Fig.~\ref{Fig4}(e), hybridization of the $\gamma$ band near the $M$ point leads to a characteristic band splitting, with the lower hybridized branch pushed below $E_{\mathrm{F}}$, thereby generating the sharp quasiparticle peak observed experimentally. This process also enhances the spectral weight near the $M$ point on the Fermi surface, consistent with the ARPES intensity distribution in Fig.~\ref{Fig3}(a). A similar hybridization scenario can be inferred for the dispersive $\varepsilon$ band near the $K$ point. Together, the schematic hybridization picture in Fig.~\ref{Fig4}(d) and Fig.~\ref{Fig4}(e) provides a unified framework for understanding the main low-temperature signatures observed by both STM/STS and ARPES.

At low temperatures, the hybridized electronic structure can be described within the framework of the periodic Anderson model (PAM)\cite{BookHeavyFermions1993,coleman2015heavyfermionskondolattice}, in which the hybridized bands are written as
\begin{equation}
E^{\pm}(k)=\frac{\epsilon_0+\epsilon_k}{2}\pm \frac{1}{2}\sqrt{(\epsilon_0-\epsilon_k)^2+4|V_k|^2},
\end{equation}
where $\epsilon_0$ denotes the renormalized energy of the localized flat band, $\epsilon_k$ represents the bare conduction-band dispersion, and $V_k$ is the momentum-dependent hybridization strength.

In the present system, the role of the localized level is not played by an atomic $f$ orbital, but by the correlation-renormalized kagome flat-band state. Within conventional Kondo lattice physics, the localized moments originate from atomic $f$ orbitals, whereas the itinerant carriers are derived from $s/p$ or $d$ electrons\cite{SchmidtNature2010,BookHeavyFermions1993,HeavyFermion2019}. In CsCr$_6$Sb$_6$, by contrast, our experiments suggest a distinct route: strong electron correlations drive a Mott-like splitting of the kagome-derived flat band that lies near $E_{\mathrm{F}}$ and is expected to be close to half filling, and the resulting occupied component supplies the localized spins that participate in low-temperature hybridization. In this sense, the localization is not atomic in origin, but instead emerges from correlated Cr 3$d$ electrons confined by kagome-lattice geometry.

\begin{figure*}[tbp]
\begin{center}
\includegraphics[width=1\textwidth,angle=0]{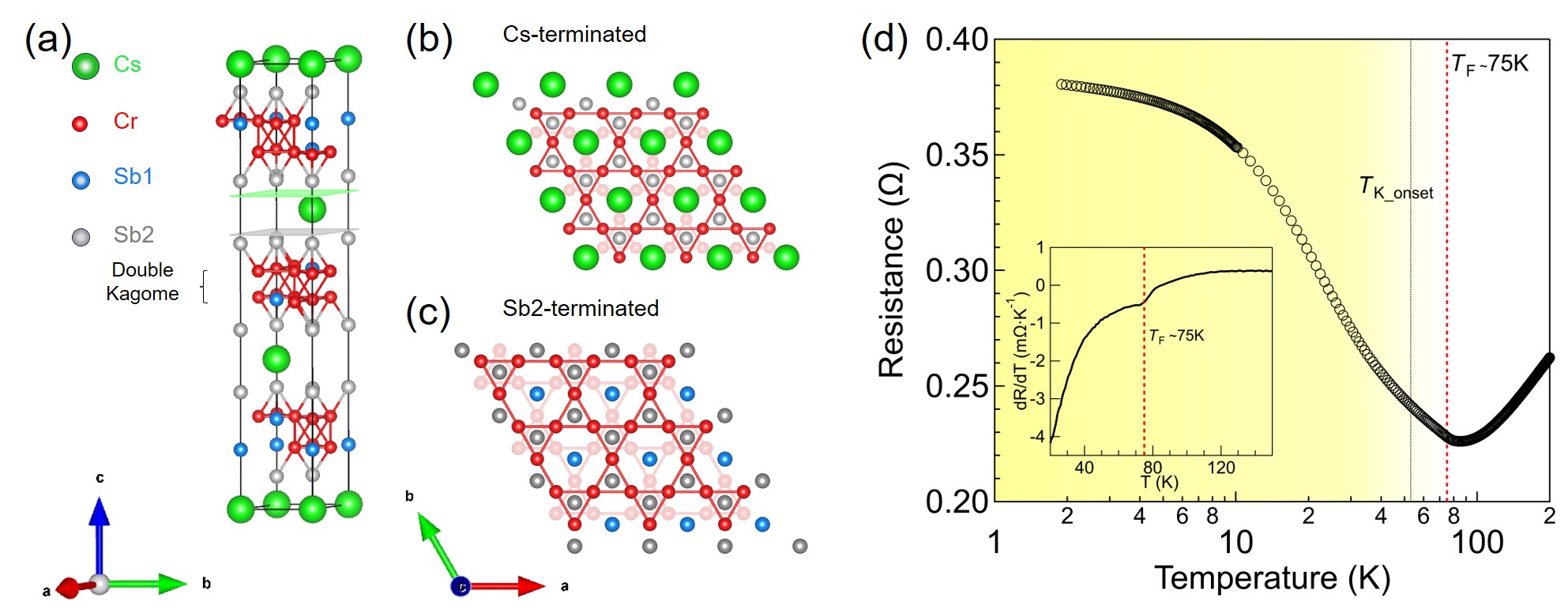}
\end{center}
\caption{{Crystal structure and transport properties of CsCr$_6$Sb$_6$.} 
(a) Side view of the crystal structure. The green and gray planes represent the Cs-terminated and Sb2-terminated surfaces, respectively. The double kagome (DK) planes are located between the Sb2 layers. 
(b) Atomic configuration of the Cs-terminated surface.
(c) Atomic model of the Sb2-terminated surface. 
(d) Temperature dependence of resistance for bulk CsCr$_6$Sb$_6$. Inset: first derivative d$R$/d$T$ showing a kink at $T_{\mathrm{F}} \sim 75$~K, indicative of frustrated magnetism. The characteristic temperature $T_{\mathrm{K\_onset}}$ is extracted from the experimental results of STM and ARPES.
}
\label{Fig1}
\end{figure*}

\begin{figure*}[tbp]
\begin{center}
\includegraphics[width=1\textwidth,angle=0]{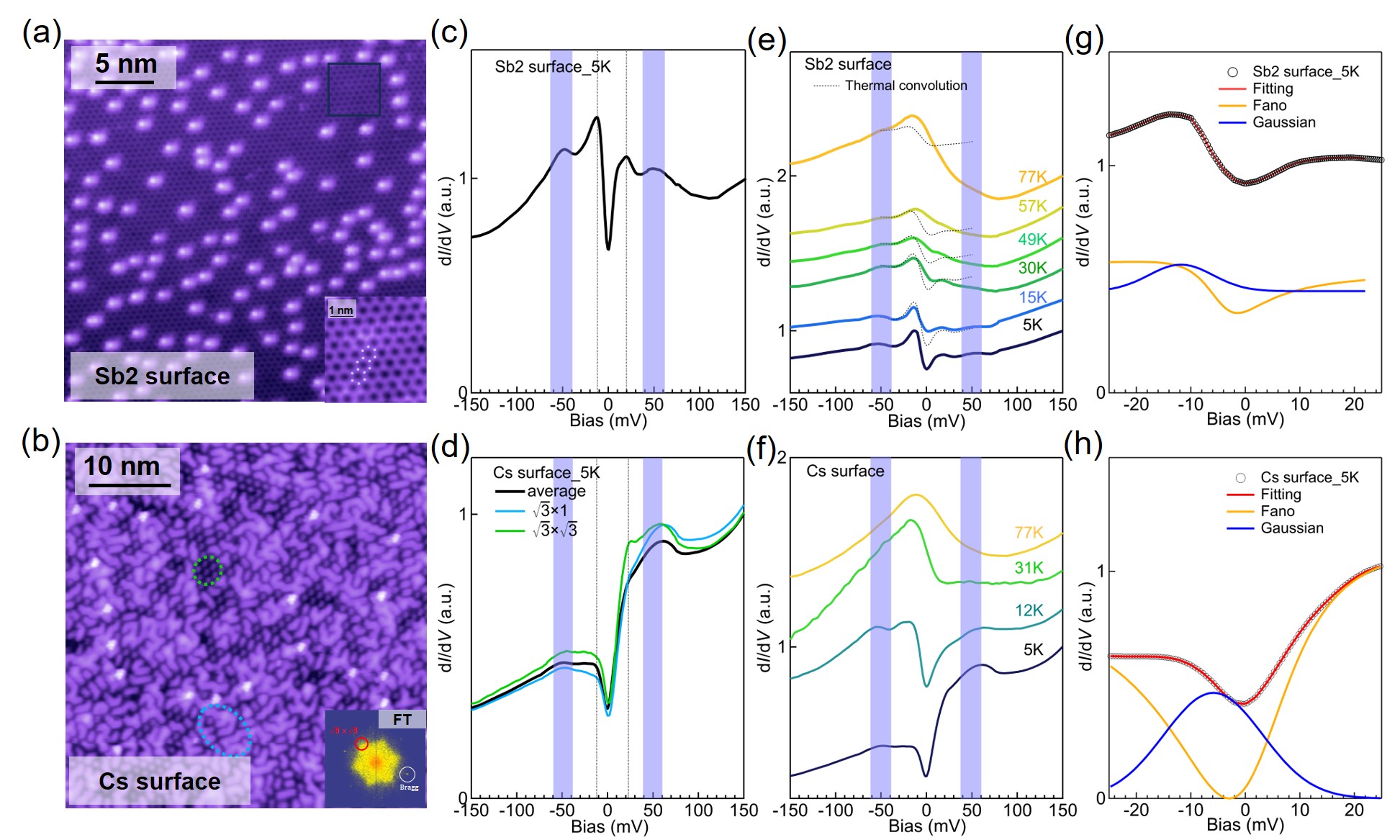}
\end{center}
\caption{{STM/STS probes Kondo and Mott physics.}
(a) Atomically resolved STM image of the Sb2-terminated surface (200 mV, 60 pA). The inset shows a magnified view of the black-boxed region, with gray circles indicating the positions of Sb2 atoms.
(b) Atomically resolved STM topographic image of the Cs-terminated surface (200 mV, 10 pA) and the corresponding Fourier transform (FT), revealing $\sqrt{3} \times \sqrt{3}$ and $\sqrt{3} \times 1$ reconstructions, highlighted by green and blue ellipses, respectively.
(c) Spatially averaged d$I$/d$V$ spectrum on the Sb2-terminated surface at 5 K. The purple shading in panels \textbf{c}-\textbf{f} highlights the $\pm 50$ mV humps. Dashed lines indicate the energy positions of peaks.
(d) Spatially averaged d$I$/d$V$ spectrum acquired on the Cs surface at 5 K (black curve). Spectra measured on two distinct Cs reconstructions are shown in green and blue. Dashed lines mark the energy positions of peaks around the Fermi level.
(e) Temperature-dependent spatially averaged d$I$/d$V$ spectra on the Sb2-terminated surface. Dotted curves are obtained by thermally broadening the 5 K spectrum using the derivative of the Fermi–Dirac distribution. The curves are offset for clarity.
(f) Temperature-dependent spatially averaged d$I$/d$V$ spectra on the Cs-terminated surface.
(g)-(h) Spatially averaged d$I$/d$V$ spectrum at 5 K (black circles) fitted with a Fano lineshape (orange) superimposed on a Gaussian peak (blue) for the Sb2-terminated surface and Cs-terminated surface, respectively.
}
\label{Fig2}
\end{figure*}

\begin{figure*}[tbp]
\begin{center}
\includegraphics[width=1\textwidth,angle=0]{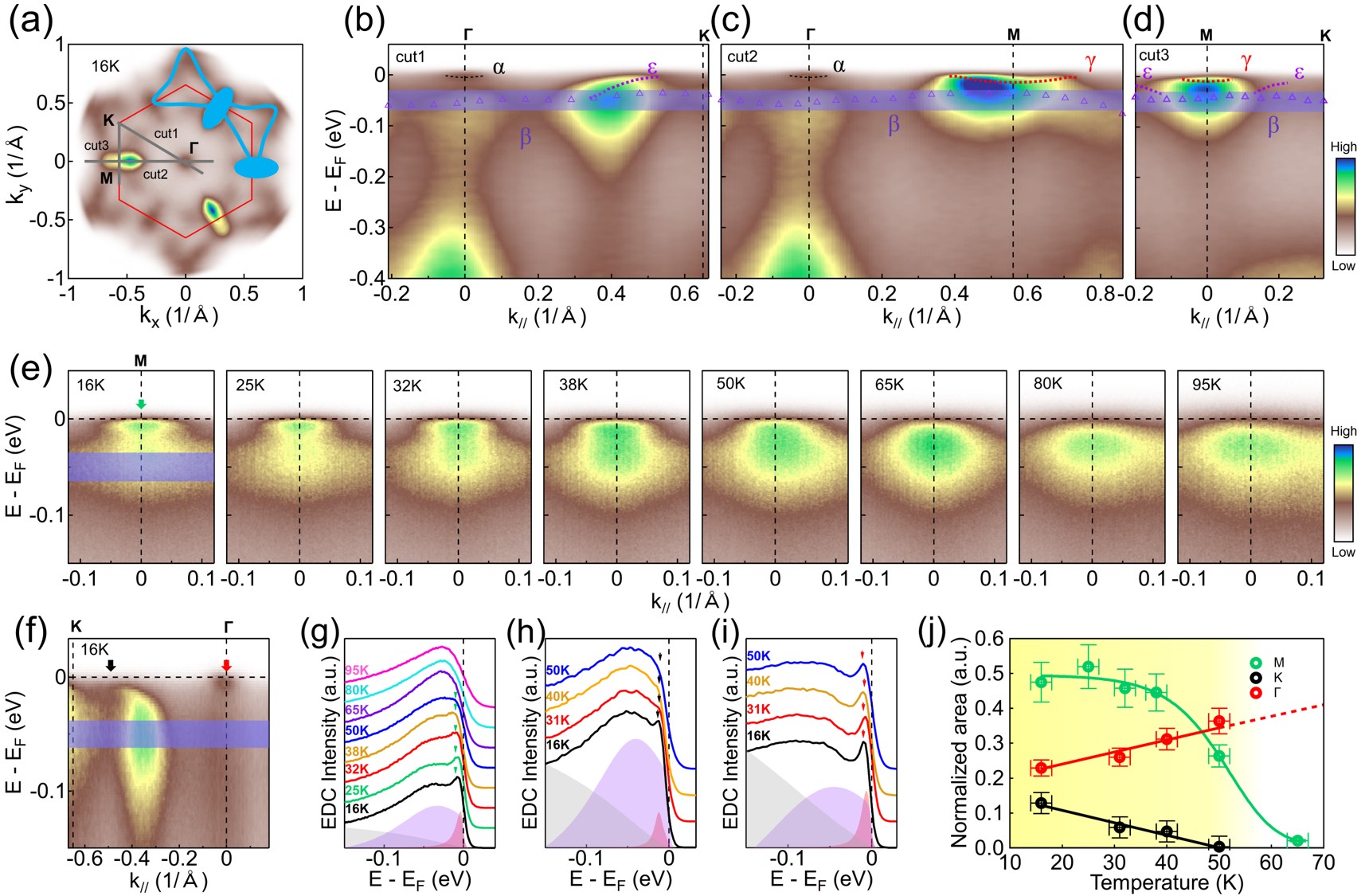}
\end{center}
\caption{{Kondo hybridization observed by ARPES in a kagome lattice.} 
(a) Fermi surface measured at 16 K using laser source
(h$\nu$ = 6.994\, eV with −69 V Bias, see Methods), obtained by integrating spectral intensity within $\pm 10$ meV energy window with respect to the $E_{\mathrm{F}}$ and symmetrized assuming three-fold rotational symmetry. Blue curves schematically depict the shape of Fermi surface.
(b)-(d) Band structures along the $\Gamma$–K, $\Gamma$–M, and M–K directions at 16 K, respectively. Four distinct bands can be distinguished, labeled $\alpha, \beta, \varepsilon, \gamma$ (highlighted by black dotted lines and triangular markers). The purple rectangular shaded areas around 50 meV below the $E_{\mathrm{F}}$ mark the location of the flat band ($\beta$ band).
(e) Temperature-dependent band structures measured at the M point along the K–M–K direction without bias. The purple shading in panel e-f marks the position of the flat band near 50 meV below $E_{\mathrm{F}}$.
(f) Low-temperature band structure measured at 16 K along the $\Gamma$-K direction with -49 V bias. 
(g)-(i) Temperature dependent energy distribution curves (EDCs) taken at the momentum positions indicated by the green arrow in panel e and the black and red arrows in panel f, corresponding to the Fermi momenta at M, near K and $\Gamma$ point, respectively. 
the EDCs were fitted after subtracting a Shirley background using two Lorentzian components multiplied by the Fermi–Dirac distribution at the corresponding temperatures. The two components represent the hump structure located at approximately $-50$ meV and the quasiparticle peak near the Fermi level, respectively. The fitted components at 16 K are shown as gray (Shirley background), purple ($-50$ meV hump), and red (quasiparticle peak) shaded areas.
(j) Temperature dependence of the normalized Lorentzian spectral weight of the quasiparticle peak extracted from the EDC fittings in panels g-i. All EDCs are normalized at 150~meV below $E_{\mathrm{F}}$ and vertically offset for clarity. Solid lines serve as guides to the eye.
}
\label{Fig3}
\end{figure*}

\begin{figure*}[tbp]
\begin{center}
\includegraphics[width=1\textwidth,angle=0]{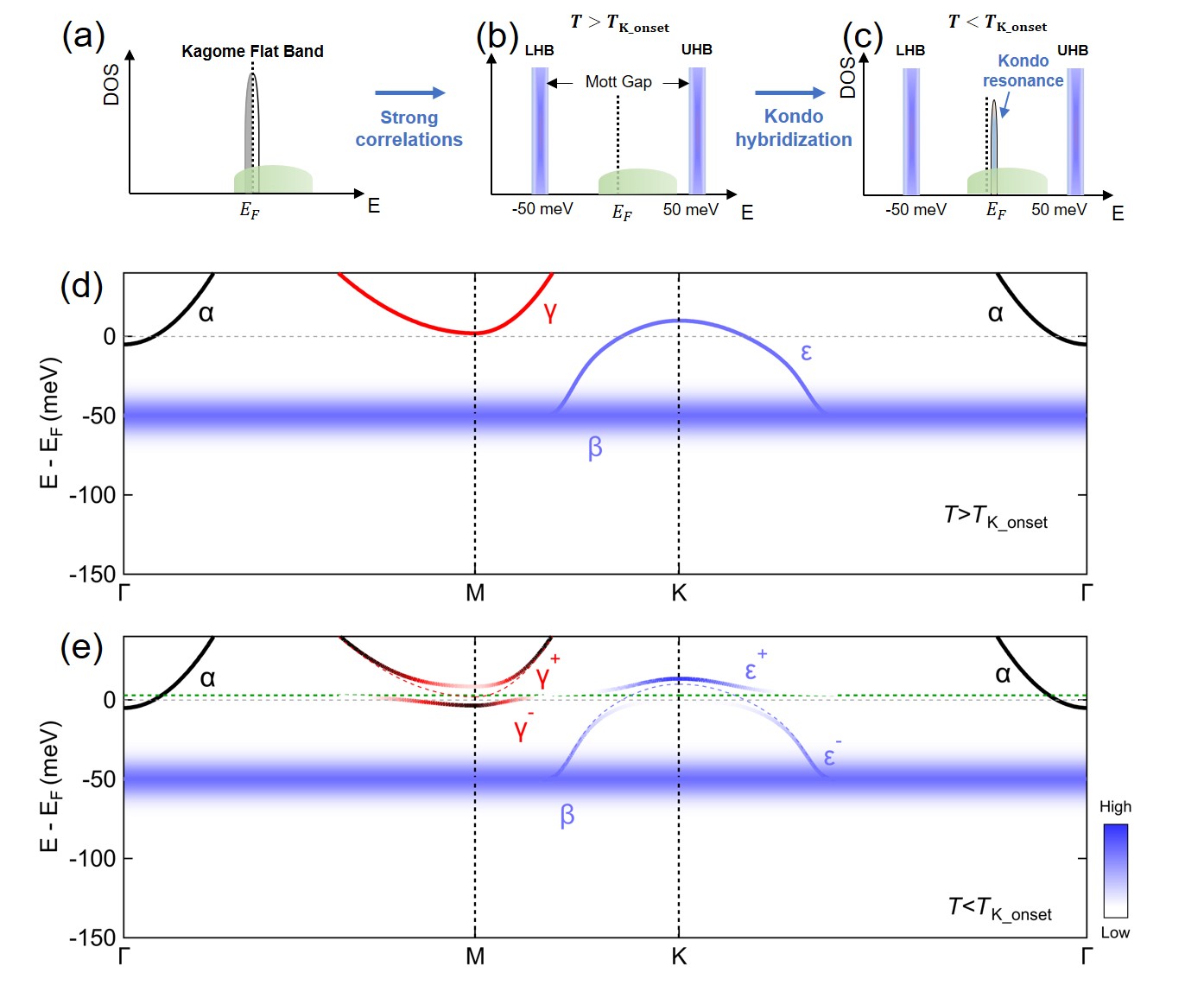}
\end{center}
\caption{{Schematic illustration of correlation-driven Kondo hybridization.} 
(a) Evolution of the density of states (DOS). In the absence of strong correlations, the half-filled kagome flat band lies at the $E_{\mathrm{F}}$. 
(b) Upon introducing strong correlations, the kagome flat band undergoes a Mott splitting into lower Hubbard band (LHB) and upper Hubbard band (UHB), opening a Mott gap.
(c) At temperature below $T_{\mathrm{K\_onset}}$, Kondo hybridization sets in, manifested by the emergence of a Kondo resonance near the $E_{\mathrm{F}}$. 
(d)-(e) Schematic band structures along high-symmetry directions based on ARPES results, illustrating momentum-dependent Kondo hybridization below $T_{\mathrm{K\_onset}}$. The conduction bands ($\varepsilon, \gamma$) hybridize with renormalized flat band (green line), forming upper and lower hybridized bands (gradient-colored lines) described within the periodic Anderson model. The hybridization strength is $V_k = 6$ meV and the renormalized flat-band level is $\epsilon_0 = 3$ meV. The $\alpha$ band remains non-hybridized. 
}
\label{Fig4}
\end{figure*}

It is also important to consider alternative mechanisms for the low-energy spectra. One possibility is that the atomic $d$ orbitals of Cr directly provide localized magnetic moments in the same sense as atomic $f$ orbitals. However, band-structure calculations indicate that all Cr 3$d$ orbitals form dispersive bands except for the kagome flat bands, making this scenario less natural. A second possibility is that magnetic impurities such as excess Cr atoms produce Kondo scattering. Yet impurity scattering alone is insufficient to form a Kondo lattice, and previous sample characterization indicates high crystalline quality with negligible impurity concentration. Another possibility is that the spectroscopic features arise from the frustrated magnetic state reported previously. This interpretation is likewise disfavored. Its characteristic temperature near 75~K, reflected in Fig.~\ref{Fig1}(d), is inconsistent with the lower spectroscopic onset temperature of 50--65~K. 
Although the nature of this magnetic state remains unclear and further investigations are required, previous studies suggest that long-range order can largely be ruled out\cite{Songboqin2025NC,XiangqiLiu2025arxiv}. Accordingly, mechanisms for gap opening driven by long-range ordering, such as ferromagnetism, antiferromagnetism, or spin-density-wave order, are unlikely to account for the observed low-energy spectra. This conclusion is further supported by the absence of any clear low-temperature ordering signature in the STM topographies shown in Fig.~\ref{Fig2}(a) and Fig.~\ref{Fig2}(b).

Finally, the transport properties provide additional context for the incomplete character of the low-temperature state. As seen in Fig.~\ref{Fig1}(d), the resistivity neither exhibits the further rapid increase expected for a fully developed Kondo insulator nor shows the downturn typical of a Fermi-liquid regime. Combined with the very low carrier concentration inferred from Hall measurements reported previously\cite{Songboqin2025NC}, this behavior suggests that the system cannot fully screen all localized moments. The resulting state may therefore be viewed as an incomplete or low-carrier-density Kondo hybridization, which naturally leaves finite spectral weight at $E_{\mathrm{F}}$ even at the lowest measured temperatures.

In summary, combined STM/STS and ARPES measurements provide direct spectroscopic evidence of Kondo hybridization in CsCr$_6$Sb$_6$ and elucidate the Mottness origin of local moments. The pronounced gap feature observed by STS and sharp quasiparticle peak observed by ARPES, together with their consistent temperature dependence, collectively demonstrate the emergence of Kondo lattice physics. In addition, the robust humps at $\pm 50$ meV in STS and the broad flat band around 50 meV below $E_{\mathrm{F}}$ in ARPES indicate a correlation-driven Mott splitting of the kagome flat band into UHB and LHB. These results establish a picture in which itinerant $d$ electrons hybridize with localized magnetic moments arising from the LHB of a Mott-split kagome flat band. Framed in this way, CsCr$_6$Sb$_6$ is not only a rare $d$-electron kagome Kondo lattice candidate but also a platform in which the microscopic connection between Mottness and Kondo hybridization can be spectroscopically tracked.

\vspace{10mm}
\noindent{\bf Methods}\vspace{3mm}\\
High-quality single crystals of CsCr$_6$Sb$_6$ were synthesized using a self-flux method\cite{Songboqin2025NC}. A mixture of Cs ingot (99.9\%, Alfa), Cr grains (99.9\%, Alfa), and Sb powder (99.999\%, Alfa) was weighed in a molar ratio of 10:3:30, loaded into an alumina crucible, and sealed in a quartz ampoule. The ampoule was heated to 1223 K and held for 24 h, then slowly cooled to 923 K at a rate of 2 K/h, followed by furnace cooling to room temperature. The resulting melt was quenched in water, yielding van der Waals–like crystals with typical dimensions of 3 mm × 3 mm × $50 \,\mu\mathrm{m}$.

STM experiments were conducted in ultrahigh-vacuum (UHV) systems. CsCr$_6$Sb$_6$ crystals were cleaved in the preparation chamber both at room temperature and low temperature (base pressure better than 5$\times$10$^{-10}$\,mbar) to expose the Cs-terminated surface as seen in  Fig.~\ref{SI_Fig2}. 
The Sb2-terminated surfaces were obtained by cleaving at around 20 K. The main STM chamber maintained a pressure better than 
2$\times$10$^{-10}$\,mbar. STM/STS measurements were performed on both Cs and Sb2 surfaces. To ensure reproducibility, STS measurements on Sb2-terminated surface were carried out on two independent STM instruments. Sample 1 was measured by a home-built STM with tungsten tips calibrated on Au(111), while sample 2 and sample 3 were measured with a Unisoku USM-1300 STM with PtIr tips calibrated on Ag(111). Tunneling conductance spectra were acquired using a standard lock-in amplifier technique.

High-resolution laser-based ARPES measurements were performed using a laboratory system equipped with a vacuum-ultraviolet laser source (h$\nu$ = 6.994\, eV) and a hemispherical electron analyzer (DA30L, Scienta-Omicron). The energy and angular resolutions were set to 1 meV and $\sim$0.3 $^\circ$, respectively\cite{Liuguodong2008RSI}. Samples are cleaved $in$ $situ$ at 15 K, under the ultra-high vacuum (pressure better than 
5$\times$10$^{-11}$\,mbar).  
A tunable bias voltage was applied to the samples to enhance momentum coverage\cite{Miao2026biasARPES}. 
The laser spot size on the sample was approximately 10 $\mu$m during measurements, which is important for searching good regions at cleaved surface.

\vspace{3mm}
\noindent {\bf Acknowledgment}\\
This work is supported by the National Key Research and Development Program of China (Grant No. 2024YFA1408301, 2021YFA1401800, 2022YFA1604200, 2022YFA1403900, 2023YFA1406002, 2023YFA1406103 and 2024YFA1408100),
the National Natural Science Foundation of China (Grant No. 12488201, 12374066, 12374154 and 12494593), CAS Superconducting Research Project (Grant No. SCZX-0101),  Quantum Science and Technology-National Science and Technology Major Project (Grant No. 2021ZD0301800) and Synergetic Extreme Condition User Facility (SECUF).

\vspace{3mm}
\noindent {\bf Author Contributions}\\
X.T.L., X.J.Z., L.Z. and X.Z. proposed and designed the research.
X.Z., X.T.L., Q.H.W., S.S.Y. and J.H.L. performed the STM/STS experiments.
X.Z. and Y.Y.X. carried out the ARPES experiment.
H.Q.M., Q.H.W. and J.H.L. contributed to the development and maintenance of home-built STM system.
T.M.M., B,L, N.C., H.C., W.P.Z., M.K.X., S.J.Z, F.F.Z., F.Y., Z.M.W., Q.J.P., H.Q.M., Z.H.Z., G.D.L., Z.Y.X., L.Z. and X.J.Z. contributed to the development and maintenance of laser-based ARPES system. 
B.Q.S., T.P.Y. and X.L.C. prepared and characterized the samples. 
W.J.L, S.L.Y. and Y.F.Y. provided theoretical understanding.
X.Z., X.T.L., L.Z. and X.J.Z. analyzed the data. 
X.Z., X.T.L., L.Z. and X.J.Z. wrote the paper.
All authors participated in the discussion and commented on the paper.

\vspace{3mm}
\noindent$^{*}$Corresponding author: Xintong Li@iphy.ac.cn, Hanqing MAO@iphy.ac.cn, ying@iphy.ac.cn, lzhao@iphy.ac.cn, XJZhou@iphy.ac.cn.
\bibliographystyle{apsrev4-1}
\bibliography{Cr166Ref}

@article{Miao2026biasARPES,
  title={Expansion of momentum space and full 2$\pi$ solid angle photoelectron collection in laser-based angle-resolved photoemission spectroscopy by applying sample bias},
  author={Miao, T. and Xu, Y. and Liang, B. and Zhu, W. and Cai, N. and Xu, M. and Wu, D. and Gu, H. and Mao, W. and Zhang, S. and Zhang, F. and Yang, F. and Wang, Z. and Peng, Q. and Xu, Z. and Zhu, Z. and Li, X. and Mao, H. and Zhao, L. and Liu, G. and Zhou, X. J.},
  journal={Review of Scientific Instruments},
  volume={97},
  number={3},
  pages={033908},
  year={2026},
  month={3},
  publisher={AIP Publishing},
  pmid={41874308}
}

@article{Fano1961PR,
  title = {Effects of Configuration Interaction on Intensities and Phase Shifts},
  author = {Fano, U.},
  journal = {Phys. Rev.},
  volume = {124},
  issue = {6},
  pages = {1866--1878},
  numpages = {0},
  year = {1961},
  month = {Dec},
  publisher = {American Physical Society},
  doi = {10.1103/PhysRev.124.1866},
  url = {https://link.aps.org/doi/10.1103/PhysRev.124.1866}
}

@article{Syozi1951,
    author = {Syôzi, Itiro},
    title = {Statistics of Kagomé Lattice},
    journal = {Progress of Theoretical Physics},
    volume = {6},
    number = {3},
    pages = {306-308},
    year = {1951},
    month = {06},
    issn = {0033-068X},
    doi = {10.1143/ptp/6.3.306},
    url = {https://doi.org/10.1143/ptp/6.3.306},

}

@article{CVScdw_PRB2021,
  title = {Electronic nature of chiral charge order in the kagome superconductor $\mathrm{Cs}{\mathrm{V}}_{3}{\mathrm{Sb}}_{5}$},
  author = {Wang, Zhiwei and Jiang, Yu-Xiao and Yin, Jia-Xin and Li, Yongkai and Wang, Guan-Yong and Huang, Hai-Li and Shao, Sen and Liu, Jinjin and Zhu, Peng and Shumiya, Nana and Hossain, Md Shafayat and Liu, Hongxiong and Shi, Youguo and Duan, Junxi and Li, Xiang and Chang, Guoqing and Dai, Pengcheng and Ye, Zijin and Xu, Gang and Wang, Yanchao and Zheng, Hao and Jia, Jinfeng and Hasan, M. Zahid and Yao, Yugui},
  journal = {Phys. Rev. B},
  volume = {104},
  issue = {7},
  pages = {075148},
  numpages = {7},
  year = {2021},
  month = {Aug},
  publisher = {American Physical Society},
  doi = {10.1103/PhysRevB.104.075148},
  url = {https://link.aps.org/doi/10.1103/PhysRevB.104.075148}
}

@article{CVScdw_PRX2021,
  title = {Three-Dimensional Charge Density Wave and Surface-Dependent Vortex-Core States in a Kagome Superconductor ${\mathrm{CsV}}_{3}{\mathrm{Sb}}_{5}$},
  author = {Liang, Zuowei and Hou, Xingyuan and Zhang, Fan and Ma, Wanru and Wu, Ping and Zhang, Zongyuan and Yu, Fanghang and Ying, J.-J. and Jiang, Kun and Shan, Lei and Wang, Zhenyu and Chen, X.-H.},
  journal = {Phys. Rev. X},
  volume = {11},
  issue = {3},
  pages = {031026},
  numpages = {8},
  year = {2021},
  month = {Aug},
  publisher = {American Physical Society},
  doi = {10.1103/PhysRevX.11.031026},
  url = {https://link.aps.org/doi/10.1103/PhysRevX.11.031026}
}

@article{Kundu2024,
  author    = {Asish K. Kundu and Xiong Huang and Eric Seewald and Ethan Ritz and Santanu Pakhira and Shuai Zhang and Dihao Sun and Simon Turkel and Sara Shabani and Turgut Yilmaz and Elio Vescovo and Cory R. Dean and David C. Johnston and Tonica Valla and Turan Birol and Dmitri N. Basov and Rafael M. Fernandes and Abhay N. Pasupathy},
  title     = {Low-energy electronic structure in the unconventional charge-ordered state of {ScV\(_6\)Sn\(_6\)}},
  journal   = {Nature Communications},
  year      = {2024},
  volume    = {15},
  number    = {1},
  pages     = {5008},
  doi       = {10.1038/s41467-024-48883-0},
  url       = {https://doi.org/10.1038/s41467-024-48883-0},
  issn      = {2041-1723}
}

@article{CVS_Nature2021,
  title   = {Roton pair density wave in a strong-coupling kagome superconductor},
  author  = {Hui Chen and Haitao Yang and Bin Hu and Zhen Zhao and Jie Yuan and Yuqing Xing and Guojian Qian and Zihao Huang and Geng Li and Yuhan Ye and Sheng Ma and Shunli Ni and Hua Zhang and Qiangwei Yin and Chunsheng Gong and Zhijun Tu and Hechang Lei and Hengxin Tan and Sen Zhou and Chengmin Shen and Xiaoli Dong and Binghai Yan and Ziqiang Wang and Hong-Jun Gao},
  journal = {Nature},
  volume  = {599},
  number  = {7884},
  pages   = {222--228},
  year    = {2021},
  doi     = {10.1038/s41586-021-03983-5},
  url     = {https://doi.org/10.1038/s41586-021-03983-5},
  issn    = {1476-4687}
}

@article{HeavyFermion2019,
  title   = {Heavy Fermion Materials and Physics},
  journal = {Acta Physica Sinica},
  volume  = {68},
  number  = {17},
  pages   = {177101},
  year    = {2019},
  issn    = {1000-3290},
  doi     = {10.7498/aps.68.20190801},
  url     = {https://wulixb.iphy.ac.cn/article/id/5e9aa6ed-8d8b-45a9-97bf-ad7a9da806b8},
  author  = {Wu Xie and Bin Shen and Yongjun Zhang and Chunyu Guo and Jiacheng Xu and Xin Lu and Huiqiu Yuan}
}

@article{ReviewRMP2020,
  title = {Colloquium: Heavy-electron quantum criticality and single-particle spectroscopy},
  author = {Kirchner, Stefan and Paschen, Silke and Chen, Qiuyun and Wirth, Steffen and Feng, Donglai and Thompson, Joe D. and Si, Qimiao},
  journal = {Rev. Mod. Phys.},
  volume = {92},
  issue = {1},
  pages = {011002},
  numpages = {19},
  year = {2020},
  month = {Mar},
  publisher = {American Physical Society},
  doi = {10.1103/RevModPhys.92.011002},
  url = {https://link.aps.org/doi/10.1103/RevModPhys.92.011002}
}

@article{RajS2005,
    title = {Angle-resolved and resonant photoemission spectroscopy on heavy-fermion superconductors {Ce}$_{2}${Co}{In}$_{8}$ and {Ce}$_{2}${Rh}{In}$_{8}$},
    author = {Raj, S. and Iida, Y. and Souma, S. and Sato, T. and Takahashi, T. and Ding, H. and Ohara, S. and Hayakawa, T. and Chen, G. F. and Sakamoto, I. and Harima, H.},
    journal = {Phys. Rev. B},
    volume = {71},
    issue = {22},
    pages = {224516},
    numpages = {8},
    year = {2005},
    month = {Jun},
    publisher = {American Physical Society},
    doi = {10.1103/PhysRevB.71.224516},
    url = {https://link.aps.org/doi/10.1103/PhysRevB.71.224516},
}

@article{Fujimori2006,
    title = {Direct observation of a quasiparticle band in {Ce}{Ir}{In}$_{5}$: An angle-resolved photoemission spectroscopy study},
    author = {Fujimori, Shinichi and Fujimori, Atsushi and Shimada, Kenya and Narimura, Takamasa and Kobayashi, Kenichi and Namatame, Hirofumi and Taniguchi, Masaki and Harima, Hisatomo and Shishido, Hiroaki and Ikeda, Shugo and Aoki, Dai and Tokiwa, Yoshifumi and Haga, Yoshinori and \ifmmode \bar{O}\else \={O}\fi{}nuki, Yoshichika},
    journal = {Phys. Rev. B},
    volume = {73},
    issue = {22},
    pages = {224517},
    numpages = {5},
    year = {2006},
    month = {Jun},
    publisher = {American Physical Society},
    doi = {10.1103/PhysRevB.73.224517},
    url = {https://link.aps.org/doi/10.1103/PhysRevB.73.224517},
}

@article{MoireKondo2023,
  author    = {Wenjin Zhao and Bowen Shen and Zui Tao and Zhongdong Han and Kaifei Kang and Kenji Watanabe and Takashi Taniguchi and Kin Fai Mak and Jie Shan},
  title     = {Gate-tunable heavy fermions in a moir{\'e} Kondo lattice},
  journal   = {Nature},
  year      = {2023},
  volume    = {616},
  number    = {7955},
  pages     = {61--65},
  doi       = {10.1038/s41586-023-05800-7},
  url       = {https://doi.org/10.1038/s41586-023-05800-7}
}

@article{vanderwaalsKondo2024,
  author    = {Kai Fan and Heng Jin and Bing Huang and Guijing Duan and Rong Yu and Zhen-Yu Liu and Hui-Nan Xia and Li-Si Liu and Yao Zhang and Tao Xie and Qiao-Yin Tang and Gang Chen and Wen-Hao Zhang and F. C. Chen and X. Luo and W. J. Lu and Y. P. Sun and Ying-Shuang Fu},
  title     = {Artificial superconducting Kondo lattice in a van der Waals heterostructure},
  journal   = {Nature Communications},
  year      = {2024},
  volume    = {15},
  number    = {1},
  pages     = {8797},
  doi       = {10.1038/s41467-024-53166-9},
  url       = {https://doi.org/10.1038/s41467-024-53166-9}
}

@article{LiVoKondoNC2012,
  author       = {Yasuhiro Shimizu and Hikaru Takeda and Moe Tanaka and Masayuki Itoh and Seiji Niitaka and Hidenori Takagi},
  title        = {An orbital-selective spin liquid in a frustrated heavy fermion spinel LiV$_2$O$_4$},
  journal      = {Nature Communications},
  year         = {2012},
  volume       = {3},
  number       = {1},
  pages        = {981},
  doi          = {10.1038/ncomms1979},
  url          = {https://doi.org/10.1038/ncomms1979},
  issn         = {2041-1723}
}

@article{Figgins2010,
    title = {Differential Conductance and Quantum Interference in Kondo Systems},
    author = {Figgins, Jeremy and Morr, Dirk K.},
    journal = {Phys. Rev. Lett.},
    volume = {104},
    issue = {18},
    pages = {187202},
    numpages = {4},
    year = {2010},
    month = {May},
    publisher = {American Physical Society},
    doi = {10.1103/PhysRevLett.104.187202},
    url = {https://link.aps.org/doi/10.1103/PhysRevLett.104.187202},
}

@article{KummerK2015,
    title = {Temperature-Independent Fermi Surface in the Kondo Lattice {YbRh}$_{2}${Si}_{2}$},
    author = {Kummer, K. and Patil, S. and Chikina, A. and G\"uttler, M. and H\"oppner, M. and Generalov, A. and Danzenb\"acher, S. and Seiro, S. and Hannaske, A. and Krellner, C. and Kucherenko, Yu. and Shi, M. and Radovic, M. and Rienks, E. and Zwicknagl, G. and Matho, K. and Allen, J. W. and Laubschat, C. and Geibel, C. and Vyalikh, D. V.},
    journal = {Phys. Rev. X},
    volume = {5},
    issue = {1},
    pages = {011028},
    numpages = {9},
    year = {2015},
    month = {Mar},
    publisher = {American Physical Society},
    doi = {10.1103/PhysRevX.5.011028},
    url = {https://link.aps.org/doi/10.1103/PhysRevX.5.011028},
}

@article{SchmidtNature2010,
  author  = {Schmidt, A. R. and Hamidian, M. H. and Wahl, P. and Meier, F. and Balatsky, A. V. and Garrett, J. D. and Williams, T. J. and Luke, G. M. and Davis, J. C.},
  title   = {Imaging the Fano lattice to `hidden order' transition in URu$_2$Si$_2$},
  journal = {Nature},
  year    = {2010},
  volume  = {465},
  number  = {7298},
  pages   = {570--576},
  doi     = {10.1038/nature09073},
  url     = {https://doi.org/10.1038/nature09073},
  issn    = {1476-4687}
}

@article{ParkWK2012,
    title = {Observation of the Hybridization Gap and Fano Resonance in the Kondo Lattice {URu}$_{2}${Si}$_{2}$},
    author = {Park, W. K. and Tobash, P. H. and Ronning, F. and Bauer, E. D. and Sarrao, J. L. and Thompson, J. D. and Greene, L. H.},
    journal = {Phys. Rev. Lett.},
    volume = {108},
    issue = {24},
    pages = {246403},
    numpages = {5},
    year = {2012},
    month = {Jun},
    publisher = {American Physical Society},
    doi = {10.1103/PhysRevLett.108.246403},
    url = {https://link.aps.org/doi/10.1103/PhysRevLett.108.246403},
}

@article{DurgaK2020SA,
    author = {Durga Khadka  and T. R. Thapaliya  and Sebastian Hurtado Parra  and Xingyue Han  and Jiajia Wen  and Ryan F. Need  and Pravin Khanal  and Weigang Wang  and Jiadong Zang  and James M. Kikkawa  and Liang Wu  and S. X. Huang },
    title = {Kondo physics in antiferromagnetic Weyl semimetal {Mn}$_{3+x}${Sn} films},
    journal = {Science Advances},
    volume = {6},
    number = {35},
    pages = {eabc1977},
    year = {2020},
    doi = {10.1126/sciadv.abc1977},
    URL = {https://www.science.org/doi/abs/10.1126/sciadv.abc1977},
}

@article{YunZhang2018,
    author = {Yun Zhang  and Haiyan Lu  and Xiegang Zhu  and Shiyong Tan  and Wei Feng  and Qin Liu  and Wen Zhang  and Qiuyun Chen  and Yi Liu  and Xuebing Luo  and Donghua Xie  and Lizhu Luo  and Zhengjun Zhang  and Xinchun Lai },
    title = {Emergence of Kondo lattice behavior in a van der Waals itinerant ferromagnet, {Fe}$_{3}${GeTe}$_{2}$},
    journal = {Science Advances},
    volume = {4},
    number = {1},
    pages = {eaao6791},
    year = {2018},
    doi = {10.1126/sciadv.aao6791},
    URL = {https://www.science.org/doi/abs/10.1126/sciadv.aao6791},
}

@article{ZahidM2020PRL,
    title = {Many-Body Resonance in a CorrelatedTopological Kagome Antiferromagnet},
    author = {Zhang, Songtian Sonia and Yin, JiaXin and Ikhlas, Muhammad and Tien, Hung-Ju and Wang, Rui and Shumiya, Nana and Chang, Guoqing and Tsirkin, Stepan S. and Shi, Youguo and Yi, Changjiang and Guguchia, Zurab and Li, Hang and Wang, Wenhong and Chang, TayRong and Wang, Ziqiang and Yang, Yifeng and Neupert, Titus and Nakatsuji, Satoru and Hasan, M. Zahid},
    journal = {Phys. Rev. Lett.},
    volume = {125},
    issue = {4},
    pages = {046401},
    numpages = {6},
    year = {2020},
    month = {Jul},
    publisher = {American Physical Society},
    doi = {10.1103/PhysRevLett.125.046401},
    url = {https://link.aps.org/doi/10.1103/PhysRevLett.125.046401},
}

@article{MottinMOF2025NSR,
    author = {Qin, Tianchen and Wang, Xingyue and Wang, Jia and Li, Xiaoyin and You, Sifan and Wang, Zihan and Zhu, Junfa and Liu, Feng and Chi, Lifeng and Pan, Minghu},
    title = {Mott state of flat bands in a two-dimensional metal−organic Kagome framework},
    journal = {National Science Review},
    pages = {nwaf574},
    year = {2025},
    month = {12},
    issn = {2095-5138},
    doi = {10.1093/nsr/nwaf574},
    url = {https://doi.org/10.1093/nsr/nwaf574},
    eprint = {https://academic.oup.com/nsr/advance-article-pdf/doi/10.1093/nsr/nwaf574/65883440/nwaf574.pdf},
}

@article{SingleKondoImpurity2002PRL,
  title = {Temperature Dependence of a Single Kondo Impurity},
  author = {Nagaoka, K. and Jamneala, T. and Grobis, M. and Crommie, M. F.},
  journal = {Phys. Rev. Lett.},
  volume = {88},
  issue = {7},
  pages = {077205},
  numpages = {4},
  year = {2002},
  month = {Feb},
  publisher = {American Physical Society},
  doi = {10.1103/PhysRevLett.88.077205},
  url = {https://link.aps.org/doi/10.1103/PhysRevLett.88.077205}
}

@article{JXY2022Nature,
    title = {Topological kagome magnets and superconductors},
    author = {Yin, Jiaxin and Lian, Biao and Hasan, M. Zahid.},
    journal = {Nature},
    volume = {612},
    pages = {647},
    numpages = {11},
    year = {2022},
    publisher = {Springer Nature},
    doi = {https://doi.org/10.1038/s41586-022-05516-0},
}

@article{WangYiLin2025PRB,
    title = {Heavy fermions in frustrated Hund's metal with portions of incipient flat bands},
    author = {Wang, Yilin},
    journal = {Phys. Rev. B},
    volume = {111},
    issue = {3},
    pages = {035127},
    numpages = {8},
    year = {2025},
    month = {Jan},
    publisher = {American Physical Society},
    doi = {10.1103/PhysRevB.111.035127},
    url = {https://link.aps.org/doi/10.1103/PhysRevB.111.035127},
}

@article{Ni3InKondoNP2024,
  author  = {Ye, Linda and Fang, Shiang and Kang, Mingu and Kaufmann, Josef and Lee, Yonghun and John, Caolan and Neves, Paul M. and Zhao, S. Y. Frank and Denlinger, Jonathan and Jozwiak, Chris and Bostwick, Aaron and Rotenberg, Eli and Kaxiras, Efthimios and Bell, David C. and Janson, Oleg and Comin, Riccardo and Checkelsky, Joseph G.},
  title   = {Hopping frustration-induced flat band and strange metallicity in a kagome metal},
  journal = {Nature Physics},
  year    = {2024},
  volume  = {20},
  number  = {4},
  pages   = {610--614},
  doi     = {10.1038/s41567-023-02360-5},
  url     = {https://doi.org/10.1038/s41567-023-02360-5},
  issn    = {1745-2481}
}

@article{Songboqin2025NC,
    title = {Realization of Kagome Kondo lattice},
    author = {Song, Boqin
    and Xie, Yuyang
    and Li, Wei-Jian
    and Liu, Hui
    and Chen, Jing
    and Tian, Shangjie
    and Zhang, Xing
    and Wang, Qinghong
    and Li, Xintong
    and Lei, Hechang
    and Zhang, Qinghua
    and Guo, Jian-gang
    and Zhao, Lin
    and Yu, Shun-Li
    and Zhou, Xingjiang
    and Chen, Xiaolong
    and Ying, Tianping},
    journal = {Nature Communications},
    volume = {16},
    issue = {1},
    pages = {5643},
    year = {2025},
    publisher = {Springer Nature},
    doi = {10.1038/s41467-025-60785-3},
}

@article{HuYong2022NC,
    title = {Rich nature of Van Hove singularities in Kagome superconductor {Cs}{V}$_{3}${Sb}$_{5}},
    author = {Hu, Y.
    and Wu, X. X.
    and Ortiz, B. R.
    and Ju, S. L.
    and Han, X. L.
    and Ma, J. Z.
    and Plumb, N. C.
    and Radovic, M.
    and Thomale, R.
    and Wilson, S. D.
    and Schnyder, A. P.
    and Shi, M.},
    journal = {Nature Communications},
    volume = {13},
    issue = {1},
    pages = {2220},
    year = {2022},
    publisher = {Springer Nature},
    doi = {https://doi.org/10.1038/s41467-022-29828-x},
}

@article{Madhavan1998Science,
    title = {Tunneling into a single magnetic atom: spectroscopic evidence of the kondo resonance},
    author = {Madhavan, V and  Chen, W and Jamneala, T and Crommie, M. F and Wingreen, N. S and Wingreen, N. S.},
    journal = {Science},
    volume = {280},
    issue = {5363},
    pages = {567},
    year = {1998},
    doi ={https://doi.org/10.1126/SCIENCE.280.5363.567},
}

@article{Aynajian2012Nature,
    title = {Visualizing heavy fermions emerging in a quantum critical Kondo lattice},
      author = {Aynajian, Pegor
    and da Silva Neto, Eduardo H.
    and Gyenis, András
    and Baumbach, Ryan E.
    and Thompson, J. D.
    and Fisk, Zachary
    and Bauer, Eric D.
    Yazdani, Ali},
    journal = {Nature},
    volume = {486},
    issue = {7042},
    pages = {201},
    year = {2012},
    publisher = {Springer Nature},
    doi = {https://doi.org/10.1038/nature11204},
}

@article{Ernst2011Nature,
  author    = {S. Ernst and S. Kirchner and C. Krellner and C. Geibel and G. Zwicknagl and F. Steglich and S. Wirth},
  title     = {Emerging local Kondo screening and spatial coherence in the heavy-fermion metal YbRh$_2$Si$_2$},
  journal   = {Nature},
  year      = {2011},
  volume    = {474},
  number    = {7351},
  pages     = {362--366},
  doi       = {10.1038/nature10148},
  url       = {https://doi.org/10.1038/nature10148}
}

@article{Ruanwei2014PRL,
    title = {Emergence of a Coherent In-Gap State in the {SmB}$_{6}$ Kondo Insulator Revealed by Scanning Tunneling Spectroscopy},
    author = {Ruan, Wei and Ye, Cun and Guo, Minghua and Chen, Fei and Chen, Xianhui and Zhang, Guang-Ming and Wang, Yayu},
    journal = {Phys. Rev. Lett.},
    volume = {112},
    issue = {13},
    pages = {136401},
    numpages = {5},
    year = {2014},
    month = {Mar},
    publisher = {American Physical Society},
    doi = {10.1103/PhysRevLett.112.136401},
    url = {https://link.aps.org/doi/10.1103/PhysRevLett.112.136401},
}

@article{Zhangyun2018PRB,
    title = {Direct observation of heavy quasiparticles in the Kondo-lattice compound $\mathrm{CeI}{\mathrm{n}}_{3}$},
    author = {Zhang, Yun and Feng, Wei and Lou, Xia and Yu, Tianlun and Zhu, Xiegang and Tan, Shiyong and Yuan, Bingkai and Liu, Yi and Lu, Haiyan and Xie, Donghua and Liu, Qin and Zhang, Wen and Luo, Xuebing and Huang, Yaobo and Luo, Lizhu and Zhang, Zhengjun and Lai, Xinchun and Chen, Qiuyun},
    journal = {Phys. Rev. B},
    volume = {97},
    issue = {4},
    pages = {045128},
    numpages = {7},
    year = {2018},
    month = {Jan},
    publisher = {American Physical Society},
    doi = {10.1103/PhysRevB.97.045128},
    url = {https://link.aps.org/doi/10.1103/PhysRevB.97.045128},
}

@article{Nb3Cl8MottPRX2023,
  title = {Discovery of a Single-Band Mott Insulator in a van der Waals Flat-Band Compound},
  author = {Gao, Shunye and Zhang, Shuai and Wang, Cuixiang and Yan, Shaohua and Han, Xin and Ji, Xuecong and Tao, Wei and Liu, Jingtong and Wang, Tiantian and Yuan, Shuaikang and Qu, Gexing and Chen, Ziyan and Zhang, Yongzhao and Huang, Jierui and Pan, Mojun and Peng, Shiyu and Hu, Yong and Li, Hang and Huang, Yaobo and Zhou, Hui and Meng, Sheng and Yang, Liu and Wang, Zhiwei and Yao, Yugui and Chen, Zhiguo and Shi, Ming and Ding, Hong and Yang, Huaixin and Jiang, Kun and Li, Yunliang and Lei, Hechang and Shi, Youguo and Weng, Hongming and Qian, Tian},
  journal = {Phys. Rev. X},
  volume = {13},
  issue = {4},
  pages = {041049},
  numpages = {9},
  year = {2023},
  month = {Dec},
  publisher = {American Physical Society},
  doi = {10.1103/PhysRevX.13.041049},
  url = {https://link.aps.org/doi/10.1103/PhysRevX.13.041049}
}

@article{Vano2021Nature,
    title = {Artificial heavy fermions in a van der Waals heterostructure},
    author = {Vaňo, Viliam
      and Amini, Mohammad
      and Ganguli, Somesh C.
      and Chen, Guangze
      and Lado, Jose L.
      and Kezilebieke, Shawulienu
      and Liljeroth, Peter},
    journal = {Nature},
    volume = {599},
    issue = {7886},
    pages = {582},
    year = {2021},
    publisher = {Springer Nature},
    doi = {10.1038/s41586-021-04021-0},
}

@Book{BookHeavyFermions1993,
    place={Cambridge}, 
    series={Cambridge Studies in Magnetism}, 
    title={The Kondo Problem to Heavy Fermions},
    publisher={Cambridge University Press}, 
    author={Hewson, Alexander Cyril}, 
    year={1993}, 
    collection={Cambridge Studies in Magnetism}
}

@misc{coleman2015heavyfermionskondolattice,
      title={Heavy Fermions and the Kondo Lattice: a 21st Century Perspective}, 
      author={Piers Coleman},
      year={2015},
      eprint={1509.05769},
      archivePrefix={arXiv},
      primaryClass={cond-mat.str-el},
      url={https://arxiv.org/abs/1509.05769}, 
}

@article{Liuguodong2008RSI,
    author = {Liu, Guodong 
    and Wang, Guiling and Zhu, Yong and Zhang, Hongbo and Zhang, Guochun and Wang, Xiaoyang and Zhou, Yong 
    and Zhang, Wentao and Liu, Haiyun and Zhao, Lin and Meng, Jianqiao and Dong, Xiaoli and Chen, Chuangtian and Xu, Zuyan and Zhou, X. J.},
    title = {Development of a vacuum ultraviolet laser-based angle-resolved photoemission system with a superhigh energy resolution better than 1meV},
    journal = {Review of Scientific Instruments},
    volume = {79},
    number = {2},
    pages = {023105},
    year = {2008},
    month = {02},
    issn = {0034-6748},
    doi = {10.1063/1.2835901},
    url = {https://doi.org/10.1063/1.2835901},
}

@Article{Zhangyun2020ReviewEng,
    title = {Application of angle-resolved photoemission spectroscopy in the study of f-electron characteristics},
    journal = {Physics},
    volume = {49},
    number = {9},
    pages = {611-619},
    year = {2020},
    issn = {0379-4148},
    doi = {10.7693/wl20200906},
    url = {https://wuli.iphy.ac.cn/cn/article/doi/10.7693/wl20200906},
    author = {Yun Zhang, Shiyong Tan, Qiuyun Chen.}
}

@misc{XiangqiLiu2025arxiv,
      title={Emergent dynamical Kondo coherence and competing magnetic order in a correlated kagome flat-band metal {Cs}{Cr}$_{6}${Sb}$_{6}$}, 
      author={Xiangqi Liu and Xuefeng Zhang and Jiachen Jiao and Renjie Zhang and Kaiwen Chen and Ying Wang and Yunguan Ye and Zhenhai Yu and Chengyu Jiang and Xia Wang and Lei Shu and Baiqing Lv and Gang Li and Yanfeng Guo},
      year={2025},
      archivePrefix={arXiv},
      primaryClass={cond-mat.str-el},
      url={https://arxiv.org/abs/2508.08580}, 
}

@article{Zhangyun2022PRB,
    title = {Kondo entanglement in the quasi-two-dimensional heavy fermion compound ${\mathrm{CeSb}}_{2}$},
  author = {Zhang, Yun and Luo, Xuebing and Feng, Wei and Tan, Shiyong and Hao, Qunqing and Zhang, Qiang and Yuan, Dengpeng and Wang, Bo and Liu, Yi and Liu, Qin and Wang, Xiyang and Luo, Lizhu and Zhu, Xiegang and Chen, Qiuyun and Lai, Xinchun},
  journal = {Phys. Rev. B},
  volume = {106},
  issue = {4},
  pages = {045133},
  numpages = {8},
  year = {2022},
  month = {Jul},
  publisher = {American Physical Society},
  doi = {10.1103/PhysRevB.106.045133},
  url = {https://link.aps.org/doi/10.1103/PhysRevB.106.045133}
}

@article{ChenQiuYun2017PRB,
  title = {Direct observation of how the heavy-fermion state develops in ${\mathrm{CeCoIn}}_{5}$},
  author = {Chen, Q. Y. and Xu, D. F. and Niu, X. H. and Jiang, J. and Peng, R. and Xu, H. C. and Wen, C. H. P. and Ding, Z. F. and Huang, K. and Shu, L. and Zhang, Y. J. and Lee, H. and Strocov, V. N. and Shi, M. and Bisti, F. and Schmitt, T. and Huang, Y. B. and Dudin, P. and Lai, X. C. and Kirchner, S. and Yuan, H. Q. and Feng, D. L.},
  journal = {Phys. Rev. B},
  volume = {96},
  issue = {4},
  pages = {045107},
  numpages = {10},
  year = {2017},
  month = {Jul},
  publisher = {American Physical Society},
  doi = {10.1103/PhysRevB.96.045107},
  url = {https://link.aps.org/doi/10.1103/PhysRevB.96.045107}
}

@article{Chenqiuyun2019PRL,
  title = {Orbital-Selective Kondo Entanglement and Antiferromagnetic Order in ${\mathrm{USb}}_{2}$},
  author = {Chen, Q. Y. and Luo, X. B. and Xie, D. H. and Li, M. L. and Ji, X. Y. and Zhou, R. and Huang, Y. B. and Zhang, W. and Feng, W. and Zhang, Y. and Huang, L. and Hao, Q. Q. and Liu, Q. and Zhu, X. G. and Liu, Y. and Zhang, P. and Lai, X. C. and Si, Q. and Tan, S. Y.},
  journal = {Phys. Rev. Lett.},
  volume = {123},
  issue = {10},
  pages = {106402},
  numpages = {5},
  year = {2019},
  month = {Sep},
  publisher = {American Physical Society},
  doi = {10.1103/PhysRevLett.123.106402},
  url = {https://link.aps.org/doi/10.1103/PhysRevLett.123.106402}
}

@article{Momentum2024CeSiI,
  title     = {Nature of the Unconventional Heavy-Fermion Kondo State in Monolayer CeSiI},
  author    = {Fumega, Adolfo O. and Lado, Jose L.},
  journal   = {Nano Letters},
  year      = {2024},
  volume    = {24},
  number    = {14},
  pages     = {4272--4278},
  publisher = {American Chemical Society},
  doi       = {10.1021/acs.nanolett.4c00619},
  url       = {https://doi.org/10.1021/acs.nanolett.4c00619},
  issn      = {1530-6984}
}

@article{Giannakis2019OrbitalSelectiveKondo,
  title   = {Orbital-selective Kondo lattice and enigmatic f electrons emerging from inside the antiferromagnetic phase of a heavy fermion},
  author  = {Giannakis, I. and Leshen, J. and Kavai, M. and Ran, S. and Kang, C. J. and Saha, S. R. and Zhao, Y. and Xu, Z. and Lynn, J. W. and Miao, L. and Wray, L. A. and Kotliar, G. and Butch, N. P. and Aynajian, P.},
  journal = {Science Advances},
  volume  = {5},
  number  = {10},
  pages   = {eaaw9061},
  year    = {2019},
  doi     = {10.1126/sciadv.aaw9061}
}

@article{ZhangW2018PRB,
  title = {ARPES/STM study of the surface terminations and $5f$-electron character in ${\mathrm{URu}}_{2}{\mathrm{Si}}_{2}$},
  author = {Zhang, W. and Lu, H. Y. and Xie, D. H. and Feng, W. and Tan, S. Y. and Liu, Y. and Zhu, X. G. and Zhang, Y. and Hao, Q. Q. and Huang, Y. B. and Lai, X. C. and Chen, Q. Y.},
  journal = {Phys. Rev. B},
  volume = {98},
  issue = {11},
  pages = {115121},
  numpages = {7},
  year = {2018},
  month = {Sep},
  publisher = {American Physical Society},
  doi = {10.1103/PhysRevB.98.115121},
  url = {https://link.aps.org/doi/10.1103/PhysRevB.98.115121}
}

@article{Chenqiuyun2018PRL,
  title = {Band Dependent Interlayer $f$-Electron Hybridization in ${\mathrm{CeRhIn}}_{5}$},
  author = {Chen, Q. Y. and Xu, D. F. and Niu, X. H. and Peng, R. and Xu, H. C. and Wen, C. H. P. and Liu, X. and Shu, L. and Tan, S. Y. and Lai, X. C. and Zhang, Y. J. and Lee, H. and Strocov, V. N. and Bisti, F. and Dudin, P. and Zhu, J.-X. and Yuan, H. Q. and Kirchner, S. and Feng, D. L.},
  journal = {Phys. Rev. Lett.},
  volume = {120},
  issue = {6},
  pages = {066403},
  numpages = {6},
  year = {2018},
  month = {Feb},
  publisher = {American Physical Society},
  doi = {10.1103/PhysRevLett.120.066403},
  url = {https://link.aps.org/doi/10.1103/PhysRevLett.120.066403}
}

@article{ShenZhiXunRMP2003,
  title = {Angle-resolved photoemission studies of the cuprate superconductors},
  author = {Damascelli, Andrea and Hussain, Zahid and Shen, Zhi-Xun},
  journal = {Rev. Mod. Phys.},
  volume = {75},
  issue = {2},
  pages = {473--541},
  numpages = {0},
  year = {2003},
  month = {Apr},
  publisher = {American Physical Society},
  doi = {10.1103/RevModPhys.75.473},
  url = {https://link.aps.org/doi/10.1103/RevModPhys.75.473}
}

@article{HuYong2023ARPES135,
    author = {Hu, Yong and Wu, Xianxin Schnyder, Andreas P. and Shi, Ming},
    title = {Electronic landscape of kagome superconductors AV3Sb5 (A = K, Rb, Cs) from angle-resolved photoemission spectroscopy},
    journal = {npj Quantum Materials},
    volume = {8},
    issue = {67},
    pages = {37-49},
    year = {2023},
    month = {11},
    issn = {0033-068X},
    doi = {10.1038/s41535-023-00599-y},
    url = {https://doi.org/10.1038/s41535-023-00599-y},
}

@article{Wangyaojia2023review,
    author = {Wang, Yaojia and Wu, Heng and McCandless, Gregory T. and Chan, Julia Y. and Ali, Mazhar N},
    title = {Quantum states and intertwining phases in kagome materials},
    journal = {Nature Reviews Physics},
    volume = {5},
    issue = {11},
    pages = {635-638},
    year = {2023},
    doi = {10.1038/s42254-023-00635-7},
    url = { https://doi.org/10.1038/s42254-023-00635-7},
}

@article{Checkelsky2024NRM,
  author  = {Checkelsky, Joseph G. and Bernevig, B. Andrei and Coleman, Piers and Si, Qimiao and Paschen, Silke},
  title   = {Flat bands, strange metals and the Kondo effect},
  journal = {Nature Reviews Materials},
  year    = {2024},
  volume  = {9},
  number  = {7},
  pages   = {509--526},
  doi     = {10.1038/s41578-023-00644-z},
  url     = {https://doi.org/10.1038/s41578-023-00644-z},
  issn    = {2058-8437}
}

@article{RbV3Sb5nanolett2022,
  title     = {Evolution of Electronic Structure in Pristine and Rb-Reconstructed Surfaces of Kagome Metal RbV$_3$Sb$_5$},
  author    = {Yu, Jiawei and Xu, Zian and Xiao, Kebin and Yuan, Yonghao and Yin, Qiangwei and Hu, Zhiqiang and Gong, Chunsheng and Guo, Yunkai and Tu, Zhijun and Tang, Peizhe and Lei, Hechang and Xue, Qi-Kun and Li, Wei},
  journal   = {Nano Letters},
  year      = {2022},
  volume    = {22},
  number    = {3},
  pages     = {918--925},
  publisher = {American Chemical Society},
  doi       = {10.1021/acs.nanolett.1c03535},
  url       = {https://doi.org/10.1021/acs.nanolett.1c03535},
  issn      = {1530-6984}
}

@article{2024arXivBerthold,
  author       = {Caiyun Chen and Jiangchang Zheng and Yuman He and Xuzhe Ying and Soumya Sankar and Luanjing Li and Yizhou Wei and Xi Dai and Hoi Chun Po and Berthold J{\"a}ck},
  title        = {Cascade of strongly correlated quantum states in a partially filled kagome flat band},
  journal      = {arXiv preprint arXiv:2409.06933},
  year         = {2024},
  month        = sep,
  archivePrefix= {arXiv},
  eprint       = {2409.06933},
  primaryClass = {cond-mat.str-el},
  doi          = {10.48550/arXiv.2409.06933}
}

@article{MITRMP1998,
  title = {Metal-insulator transitions},
  author = {Imada, Masatoshi and Fujimori, Atsushi and Tokura, Yoshinori},
  journal = {Rev. Mod. Phys.},
  volume = {70},
  issue = {4},
  pages = {1039--1263},
  numpages = {0},
  year = {1998},
  month = {Oct},
  publisher = {American Physical Society},
  doi = {10.1103/RevModPhys.70.1039},
  url = {https://link.aps.org/doi/10.1103/RevModPhys.70.1039}
}

@article{Brandow1977,
author = {B.H. Brandow},
title = {Electronic structure of Mott insulators},
journal = {Advances in Physics},
volume = {26},
number = {5},
pages = {651--808},
year = {1977},
publisher = {Taylor \& Francis},
doi = {10.1080/00018737700101443},
URL ={https://doi.org/10.1080/00018737700101443},
eprint = {https://doi.org/10.1080/00018737700101443}
}
\vspace{3mm}
\newpage




\newpage

\vspace{3mm}

\newpage
\clearpage
\setcounter{figure}{0}
\renewcommand{\thefigure}{S\arabic{figure}}
\makeatletter
\renewcommand{\theHfigure}{S\arabic{figure}} 
\makeatother
\noindent{\large\bf  Supplementary Materials}\vspace{3mm}\\
\hspace*{6mm}
\noindent\textbf{1. Ubiquity of the gap feature and $\pm$50 mV humps in CsCr$_6$Sb$_6$ single crystals}\vspace{3mm}\\
\hspace*{6mm} 
To verify the reproducibility of the gap feature and the $\pm$50 mV humps, we performed STM/STS measurements using different instruments and tip calibration procedures (see Methods). 
As shown in Fig.~\ref{SI_Fig1}, sample 1 was measured with a PtIr tip calibrated on an Ag(111) with a Unisoku USM-1300 STM, while samples 2 and 3 were measured with tungsten tips calibrated on an Au(111) using a home-built STM. 
Fig.~\ref{SI_Fig2}(a) shows three types of Cs surface reconstructions, indicative of different Cs-atom coverages. The $\sqrt{3} \times \sqrt{3}$ and $\sqrt{3} \times \ 1$ reconstructions have been observed in several samples, with representative topographies and STS spectra shown in Fig.~\ref{SI_Fig2}(b) and Fig.~\ref{SI_Fig2}(c), respectively. 
Spatially resolved d$I$/d$V$ spectra acquired from different regions of sample 2 are presented in Fig.~\ref{SI_Fig2}(d). 
Fig.~\ref{SI_Fig3} compares STS spectra measured on the clean Sb2 surface and at Cs impurities. Although spectra taken at Cs impurities exhibit an enhanced positive-bias background, the gap feature and the characteristic $\pm$50 mV humps remain clearly visible.
These results demonstrate that the gap feature around the $E_{\mathrm{F}}$ and the $\pm$50 mV humps are robust against variations in Cs surface reconstructions and are ubiquitous in this system.
Fig.~\ref{SI_Fig5}(a)-(b) show the topographic image of a step edge on the Sb2 surface measured at 5 K. The height of the step edge is around 6.2 nm, which is not an integer multiple of the lattice constant (3.45 nm). This indicates the upper and lower terraces correspond to different one-third unit-cell terminations possibly, as seen in the graph on the left of Fig.~\ref{SI_Fig5}(a). The d$I$/d$V$ spectra on both terraces are displayed in Fig.~\ref{SI_Fig5}(c).
Despite differing background contributions, the d$I$/d$V$ spectra from both terraces consistently show the same gap and robust ±50 mV humps, underscoring their robustness and universality.

\vspace{10mm}
\noindent\textbf{2. Temperature-evolution of the electronic structure probed by STS and ARPES}\vspace{3mm}\\
\hspace*{6mm} 
Key manifestations of Kondo hybridization in this system are reflected in the temperature evolution of the electronic structure, which are consistently observed across independent experiments, as illustrated in Fig.~\ref{SI_Fig7} and Fig.~\ref{SI_Fig9}.
As shown in Fig.~\ref{SI_Fig7}, the tunneling spectra on Sb2 surface display a gradual closure of the gap upon warming toward $T_{\mathrm{K\_onset}}$, consistent with results in Fig.~\ref{Fig2}(e). 
Complementary ARPES measurements (Fig.~\ref{SI_Fig9}) reveal a pronounced spectral weight near the $E_{\mathrm{F}}$ at the M point along $\Gamma$-K direction, whose temperature-dependent EDCs follow the same evolution trend as those presented in Fig.~\ref{Fig3}(g).
Together, these reproducible STS and ARPES observations provide further support for the presence of Kondo hybridization and confirm the Kondo lattice physics in CsCr$_6$Sb$_6$.

\begin{figure*}[tbp]
\begin{center}
\includegraphics[width=0.8\textwidth,angle=0]{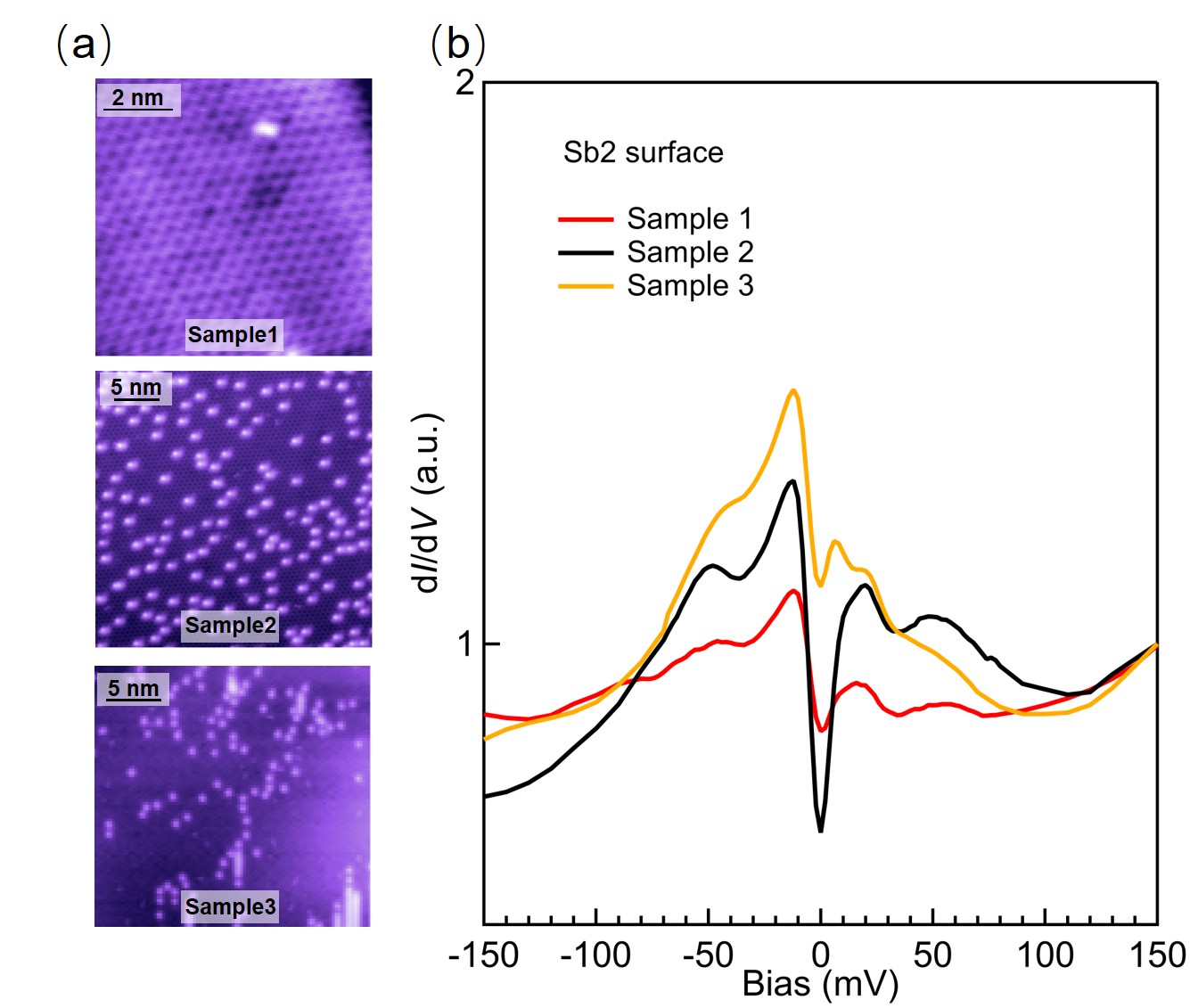}
\end{center}
\caption{{\bf Reproducibility of STM topographies and spatially averaged d$I$/d$V$ spectra on the Sb2 surface.} 
(a) STM topographies acquired from three different samples, with occasional bright Cs clusters.
(b) Corresponding spatially averaged d$I$/d$V$ spectra measured under different experimental conditions on the clean Sb2 surface. 
}
 \label{SI_Fig1}
\end{figure*}

\begin{figure*}[tbp]
\begin{center}
\includegraphics[width=1\textwidth,angle=0]{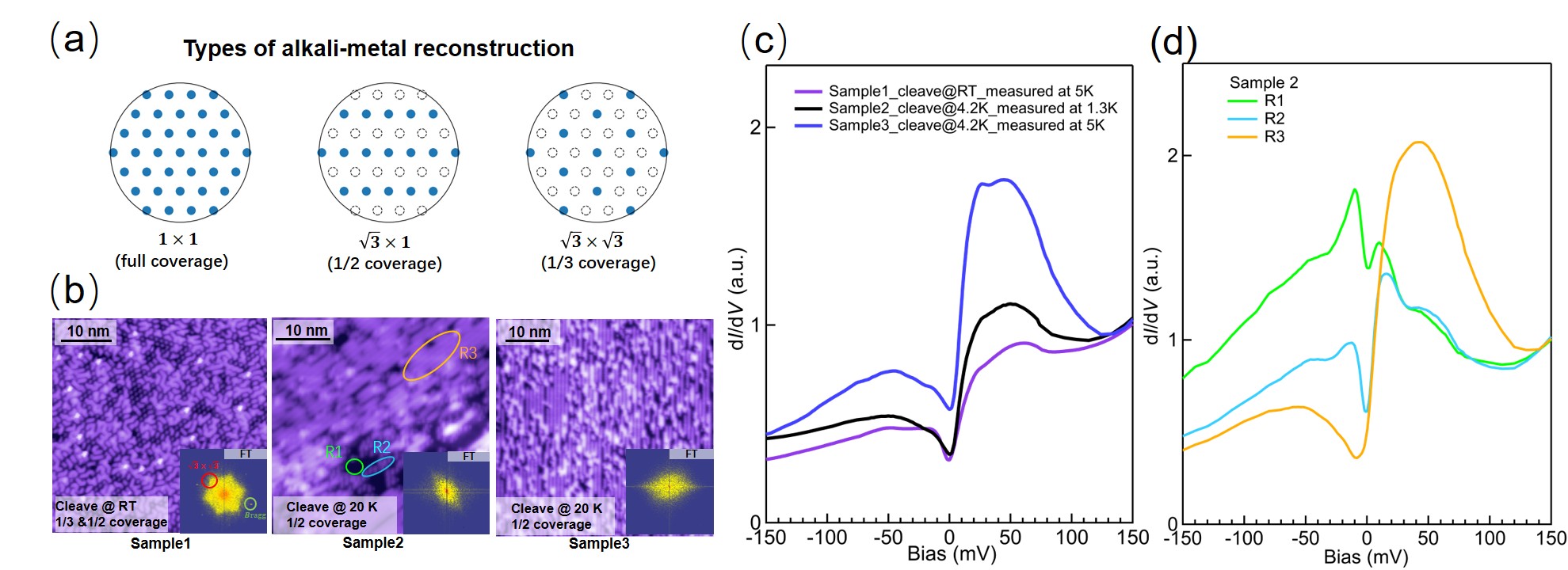}
\end{center}
\caption{{\bf Cs surface reconstructions and corresponding low-temperature STS spectra.} 
(a) Representative Cs surface reconstruction patterns corresponding to different Cs coverages on the cleaved surface.
(b) Cs reconstructions obtained from different samples at low temperature, with black characters indicating the corresponding cleaving conditions. Insets show the corresponding Fourier transform (FT) patterns of each topography.
(c) Spatially averaged STS spectra corresponding to the Cs reconstructions shown in panel (b). 
(d) The STS spectra acquired from different regions of sample 2.}
\label{SI_Fig2}
\end{figure*}

\begin{figure*}[tbp]
\begin{center}
\includegraphics[width=1\textwidth,angle=0]{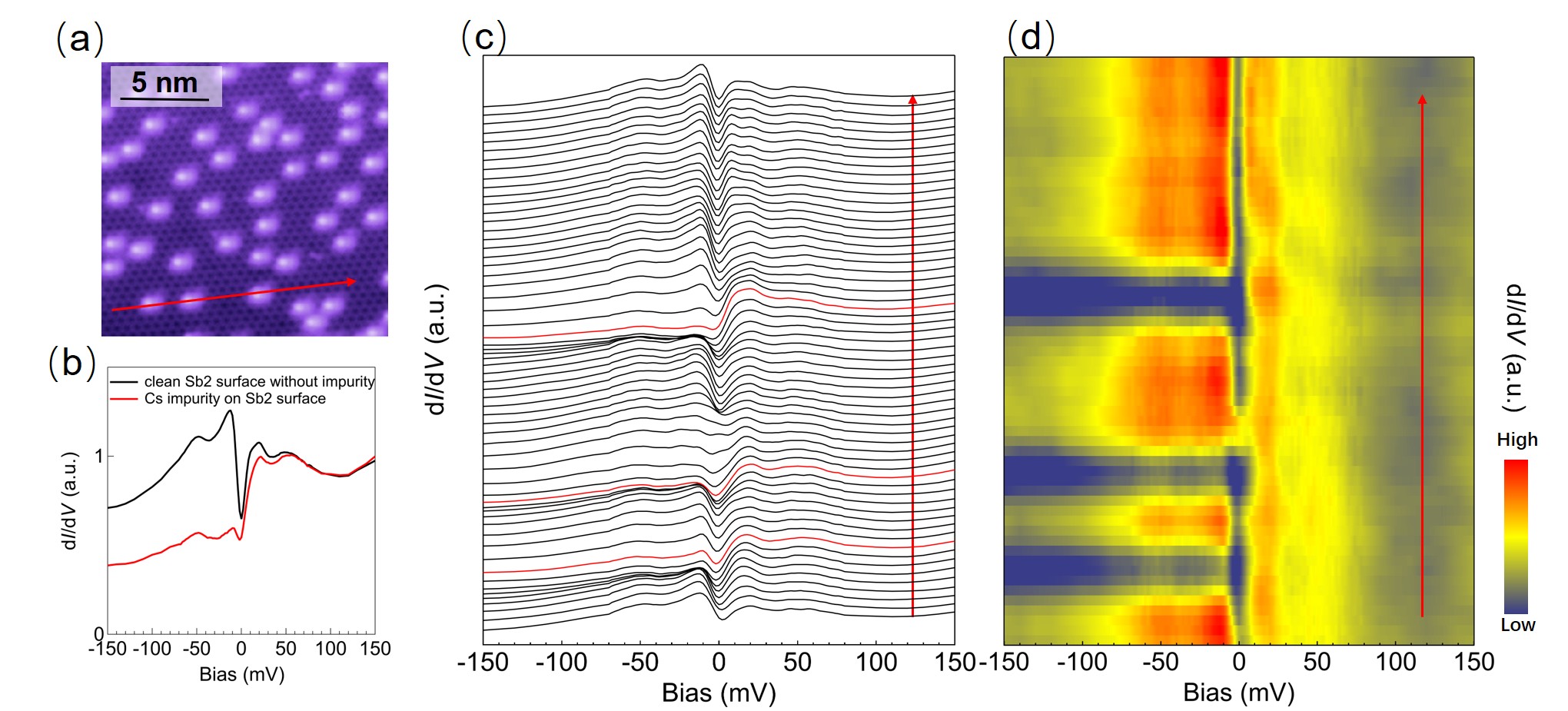}
\end{center}
\caption{{\bf STS spectra on the clean Sb2 surface and at Cs impurities.}
(a) The topography of Sb2 surface with bright Cs impurities. 
(b)  d$I$/d$V$ spectra measured on the clean Sb2 surface (black curve) and at the Cs impurities (red curve).
(c) d$I$/d$V$ spectra acquired along the line cut marked by red line with arrow in panel \textbf{a}.
(d) Color map of the spatially resolved d$I$/d$V$ spectra across the line cut in panel \textbf{a}.}
\label{SI_Fig3}
\end{figure*}

\vspace{3mm}
\begin{figure*}[tbp]
\begin{center}
\includegraphics[width=0.75\textwidth,angle=0]{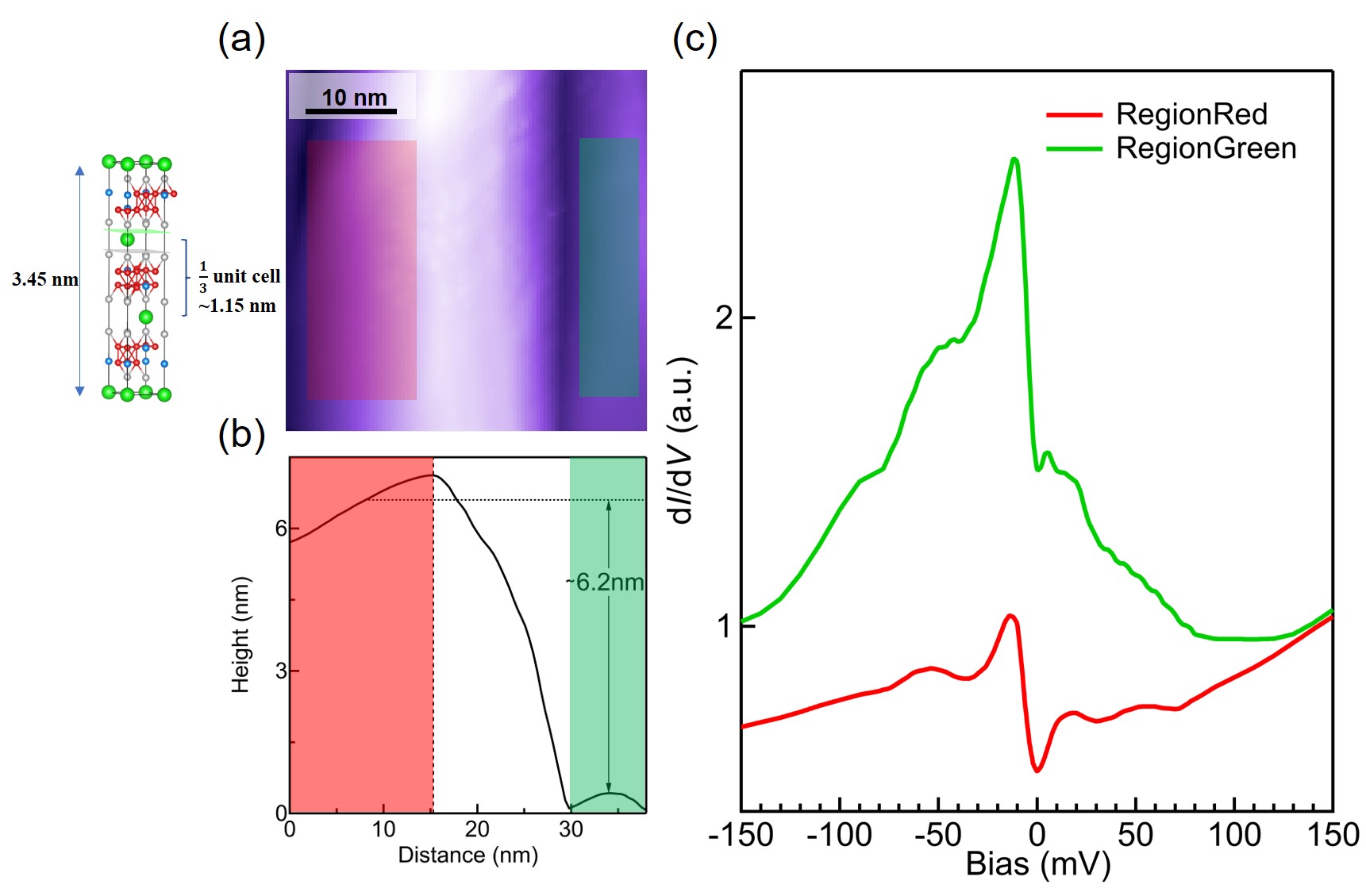}
\end{center}
\caption{{\bf Topography and tunneling conductance spectra of a step edge on the Sb2 surface at 5 K.} 
(a) Topography of a step edge of Sb2 surface. The graph on the left displays the crystal structure of CsCr$_6$Sb$_6$.
(b) The corresponding height profile of the step edge in panel (a). 
(c) d$I$/d$V$ spectra measured on the upper (red area) and lower (green area) terraces, showing different background intensities while preserving the gap feature and the $\pm$50 mV humps.}
\label{SI_Fig4}
\end{figure*}

\begin{figure*}[tbp]
\begin{center}
\includegraphics[width=0.6\textwidth,angle=0]{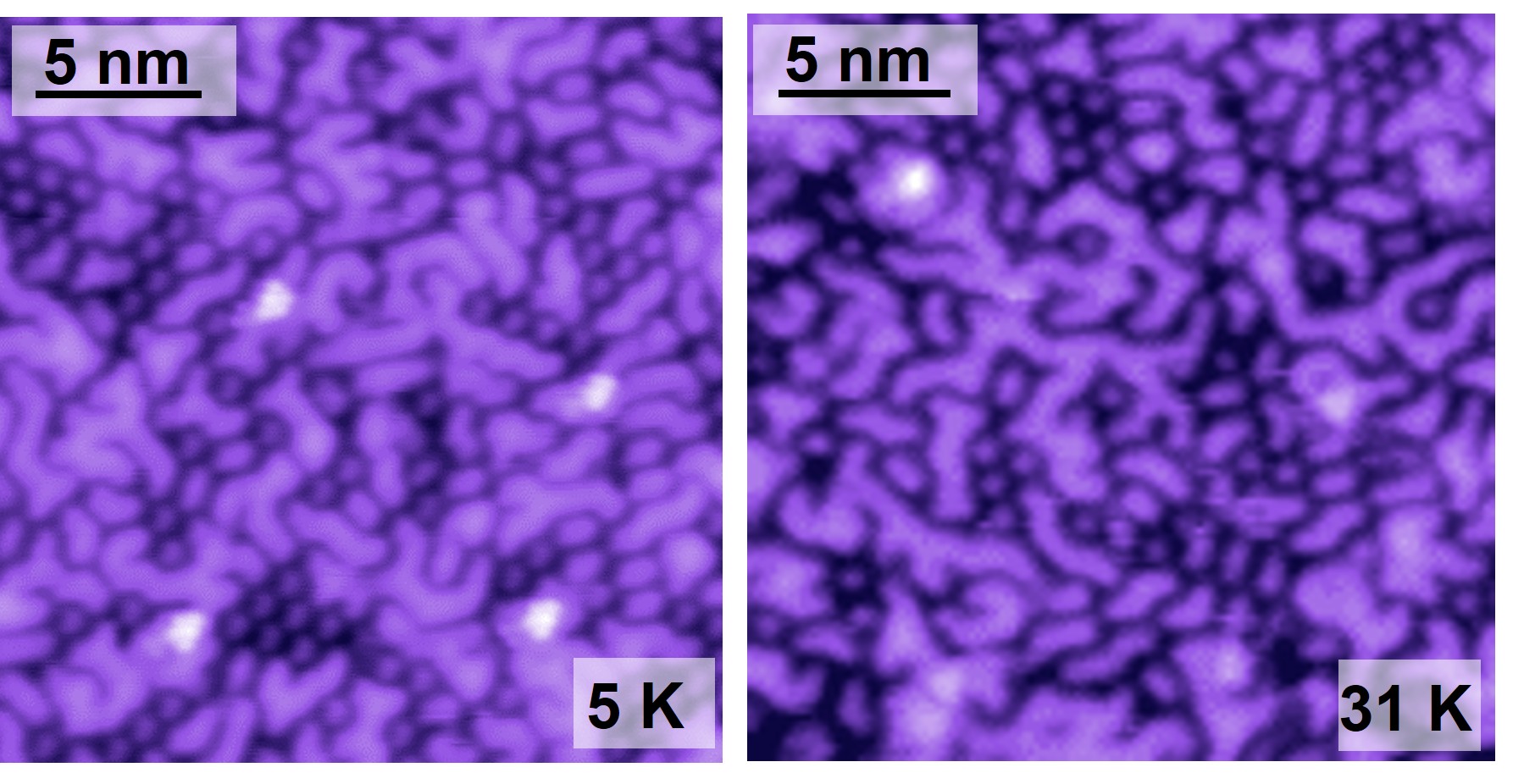}
\end{center}
\caption{{\bf Temperature stability of Cs surface reconstructions between 5 K and 31 K.}}
\label{SI_Fig5}
\end{figure*}

\begin{figure*}[tbp]
\begin{center}
\includegraphics[width=0.8\textwidth,angle=0]{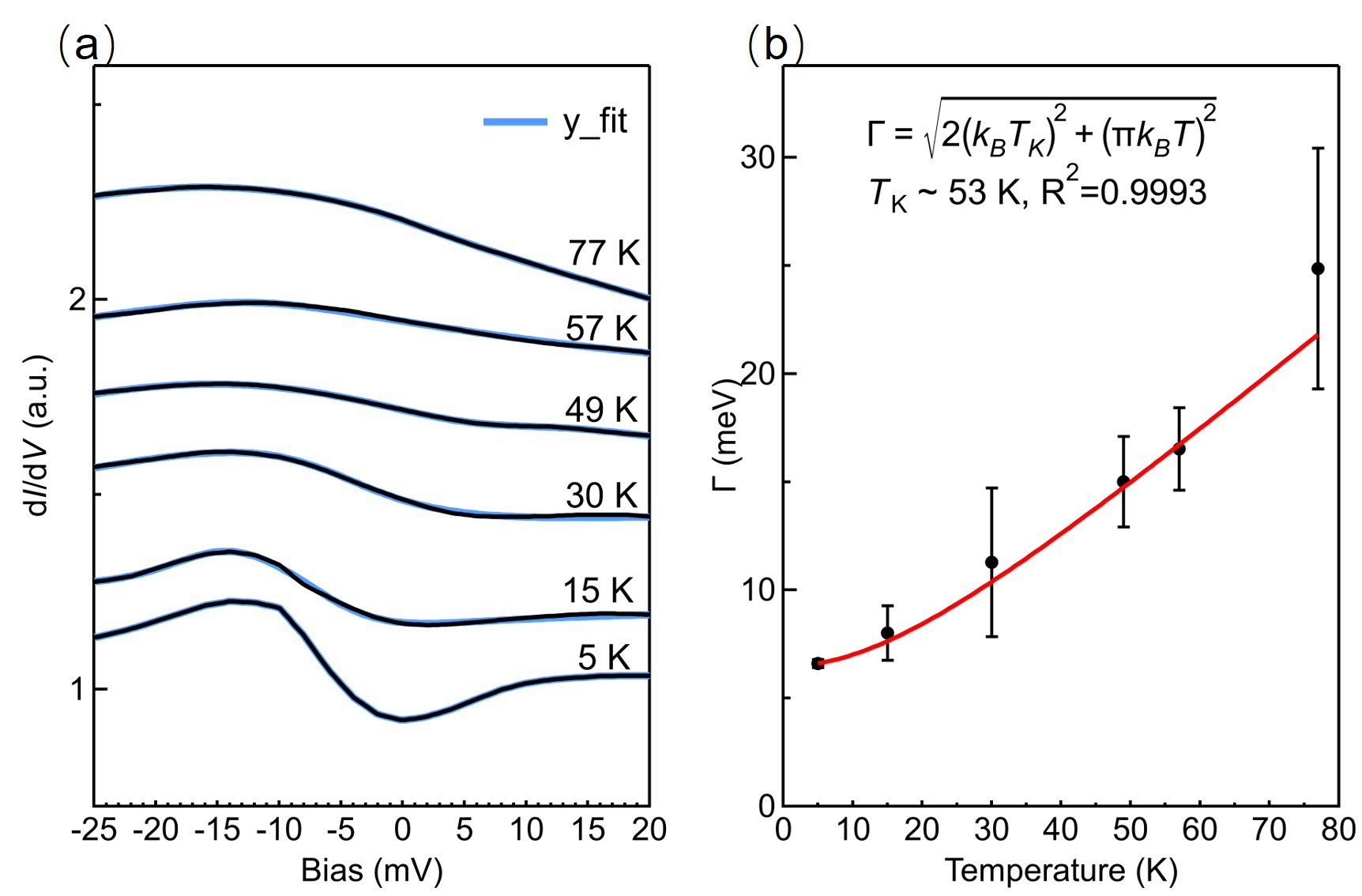}\end{center}
\caption{{\bf Temperature-dependent fitting of STS spectra on the Sb2 surface.} 
(a) STS spectra at different temperatures fitted using a Fano plus Gaussian function, as shown in Fig.~\ref{Fig2}(g). Black curves denote the raw data and blue curves the fitted results.
(b) Temperature dependence of the extracted Fano resonance width. The red line denotes the expected temperature dependence for a single Kondo impurity.
}
 \label{SI_Fig6}
\end{figure*}

\begin{figure*}[tbp]
\begin{center}
\includegraphics[width=0.4\textwidth,angle=0]{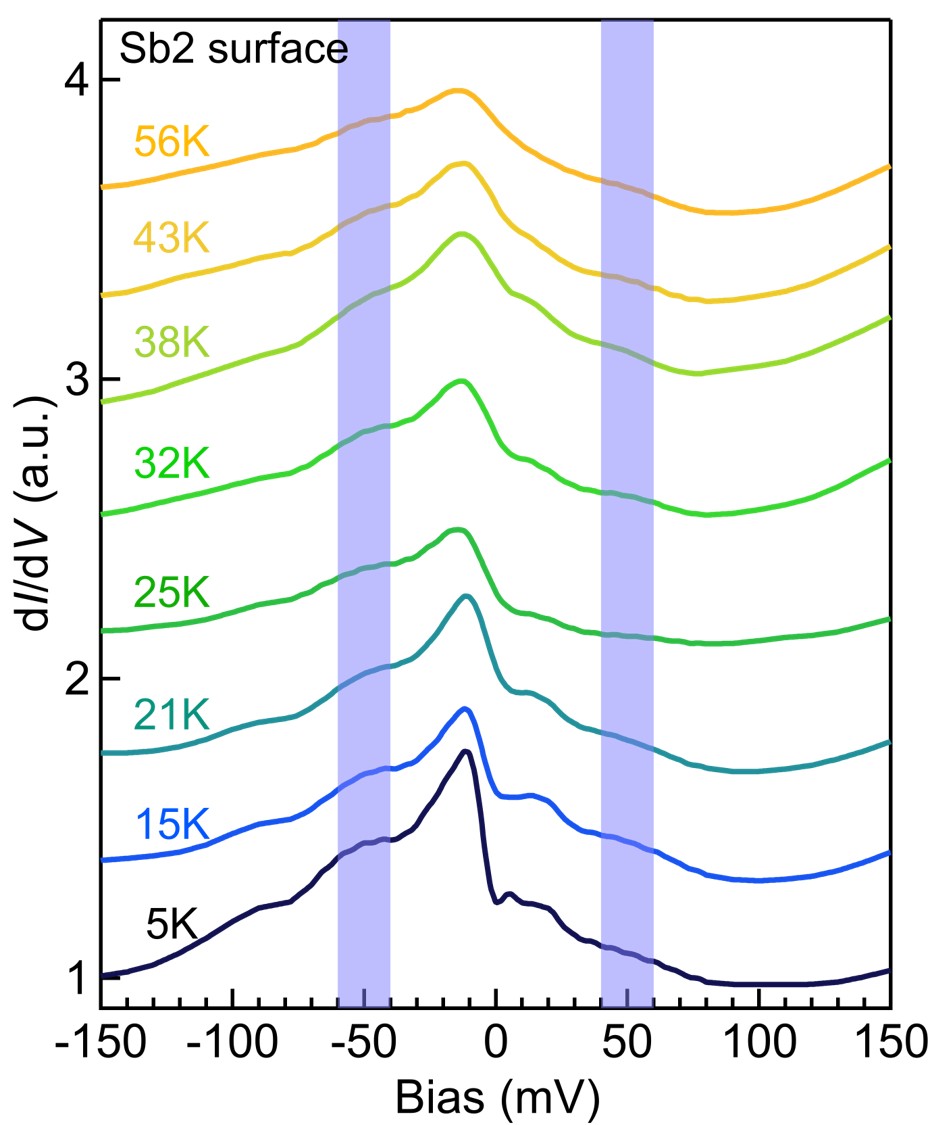}
\end{center}
\caption{{\bf Temperature-dependent STS spectra on the Sb2 surface from another experiment.} 
STS spectra acquired using a PtIr tip calibrated on Ag(111), showing a temperature evolution of the gap feature and $\pm$50 mV humps consistent with Fig.~\ref{Fig2}(f).}
\label{SI_Fig7}
\end{figure*}

\vspace{3mm}
\begin{figure*}[tbp]
\begin{center}
\includegraphics[width=0.9\textwidth,angle=0]{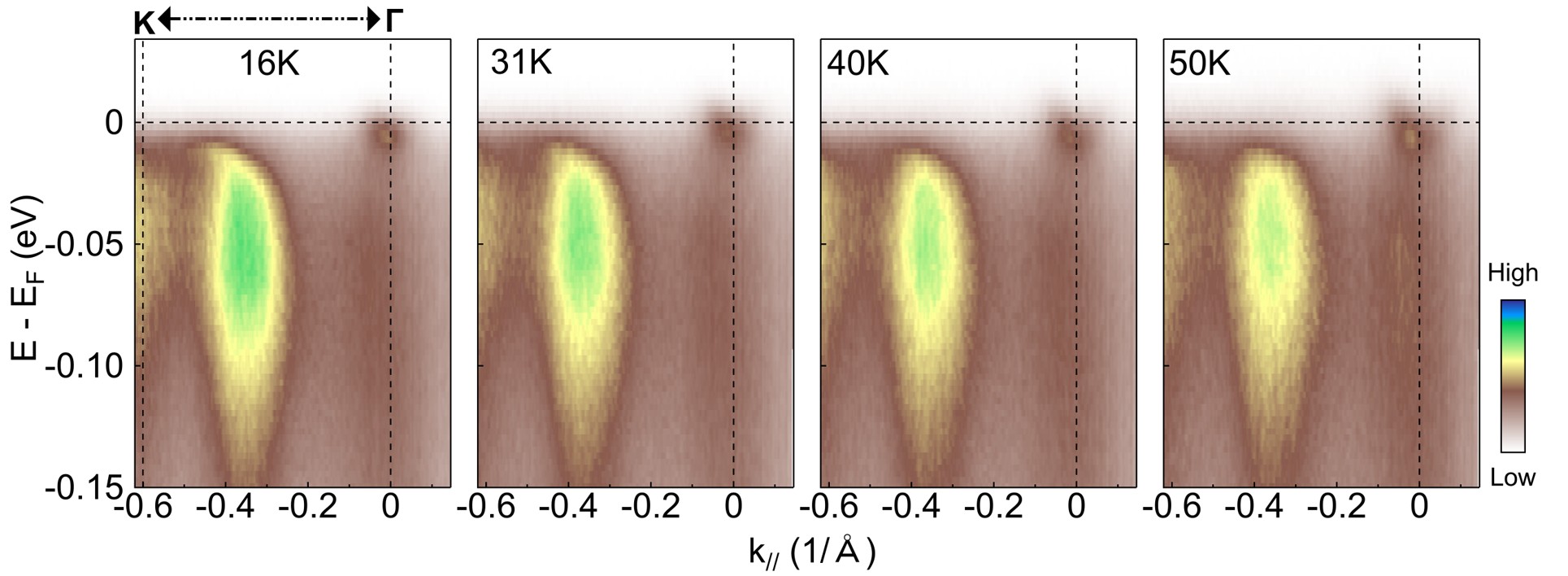}
\end{center}
\caption{{\bf Temperature-dependent band structure along the $\Gamma$-K direction (with -49 V Bias).}}
\label{SI_Fig8}
\end{figure*}

\begin{figure*}[tbp]
\begin{center}
\includegraphics[width=1\textwidth,angle=0]{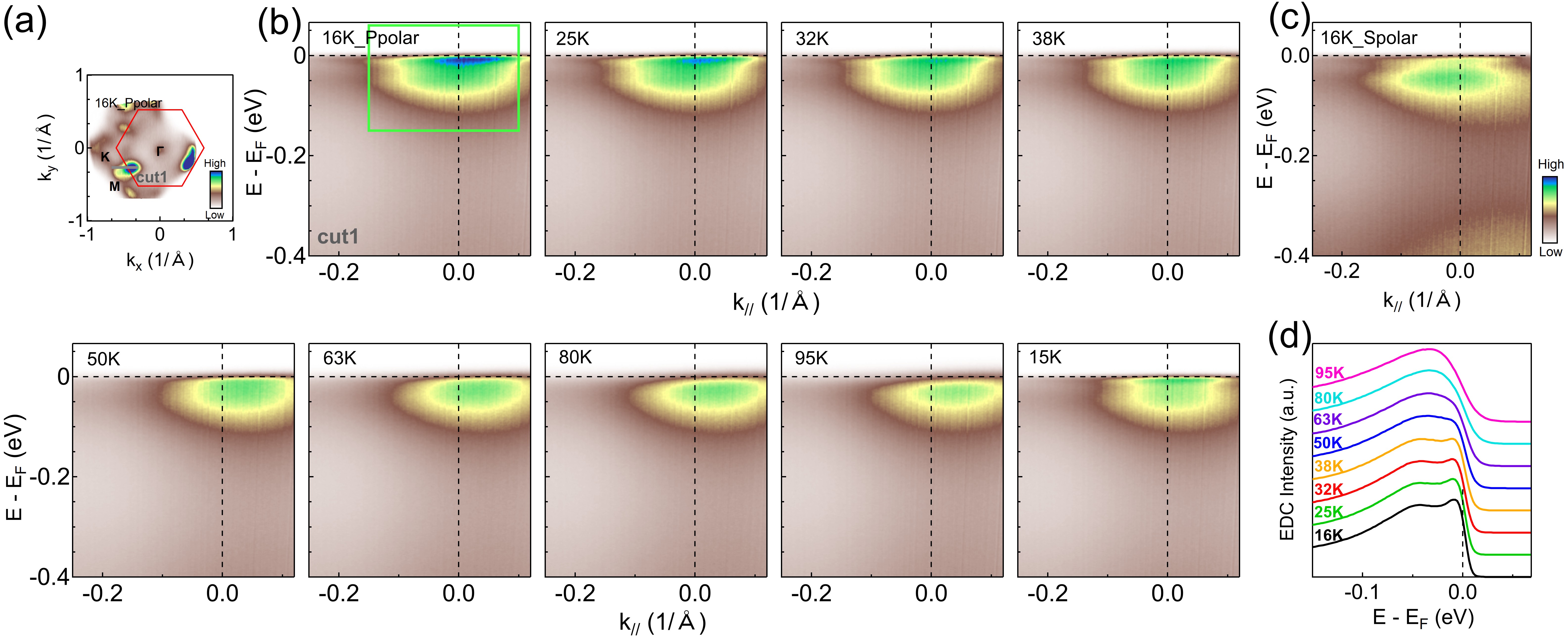}
\end{center}
\caption{{\bf Reproducibility of ARPES quasiparticle features near $E_{\mathrm{F}}$.}
(a) Fermi surface mapping of CsCr$_6$Sb$_6$ measured at 15 K. It is obtained by integrating the spectral intensity within 10 meV with respect to the $E_{\mathrm{F}}$.
(b) Temperature-dependent band structure at the M point along $\Gamma$-K direction (h$\nu$ = 6.994\, eV ,without bias). A repeated measurement at 15 K after warming to 95 K yields identical results, confirming data reliability.
(c) The same cut measured at 16 K using different light polarization (S-polarized, without bias), showing persistent spectral weight near $E_{\mathrm{F}}$.
(d) EDCs integrated over the momentum window marked by the green rectangle in panel \textbf{a}.}
\label{SI_Fig9}
\end{figure*}

\end{document}